\documentclass[preprintnumbers,amsmath,amssymb]{revtex4}

\usepackage{graphicx,subfigure}
\usepackage{dcolumn}
\usepackage{bm}
\usepackage{amsthm}
\DeclareMathOperator{\sech}{sech}

\begin{document}

\title{Classification of the classical $SL(2,\mathbb{R})$ gauge transformations in the rigid body}

\author{Manuel de la Cruz}
\email{fisikito@gmail.com}

\author{N\'{e}stor Gaspar}
\email{nex3t.gr@gmail.com}

\author{Lidia Jim\'{e}nez}
\email{lidia@xanum.uam.mx}

\author{Rom\'{a}n Linares} 
\email{lirr@xanum.uam.mx}

\affiliation{Departamento de F\'{\i}sica, Universidad Aut\'onoma Metropolitana Iztapalapa,\\
San Rafael Atlixco 186, C.P. 09340, M\'exico D.F., M\'exico,
}


\begin{abstract}
In this paper we revisit the classification of the gauge transformations in the Euler top system using the generalized 
classical Hamiltonian dynamics of Nambu. In this framework the Euler equations of motion are bi-Hamiltonian and 
$SL(2, \mathbb{R})$ linear combinations of the two Hamiltonians leave the equations of motion invariant, although 
belonging to inequivalent Lie-Poisson structures. Here we give the explicit form of the Hamiltonian vector fields 
associated to the components of the angular momentum for every single Lie-Poisson structure including both the 
asymmetric rigid bodies and its symmetric limits. We also give a detailed classification of the different Lie-Poisson 
structures recovering all the ones reported previously in the literature.
\end{abstract}

\maketitle

\section{Introduction}

Taking the Liouville theorem as a guiding principle, in \cite{Nambu:1973qe} Yoichiro Nambu proposed a 
generalization of the classical Hamiltonian dynamics by supersede the usual two dimensional phase space with 
an $n$-dimensional one. The dynamics in the new phase space was formulated via an $n$-linear fully 
antisymmetric bracket (Nambu bracket) with two or more ``Hamiltonians'' instead of the standard Poisson bracket 
with one Hamiltonian. As an example Nambu applied his formalism to a well understood system in both classical 
and quantum mechanics, the free rigid body or Eulerian top for short. For this system the phase space is 
three-dimensional  and is generated by the components of the angular momentum $\vec{L}$ in the body-fixed 
frame, the two Hamiltonians or Casimir functions are given by the total kinetic energy $E$ and the square of angular 
momentum $L^2 \equiv G$, while the Euler equations of motion are written in terms of the Nambu bracket as  
$\dot{L}_i=\{ L_i, E, G \}= \nabla L_i \cdot (\nabla E \times \nabla G)$. In this framework the problem is manifestly 
invariant under linear {\it canonical transformations} on the triplet $\vec{L} \rightarrow \vec{L}'$ which belong to the 
group $SL(3,\mathbb{R})$, and under $SL(2,\mathbb{R})$ linear combinations of the constants of motion  
$E$ and $G$, which were referred originally as {\it gauge transformations} by Nambu, nomenclature we use 
throughout this paper. It is important to note that these two sets of symmetries are not independent, in fact the 
gauge transformations are a subset of the canonical transformations \cite{Nambu:1973qe}.
Working explicitly with the symmetries in the Nambu classical dynamics is the base of its power. It has 
been recognized that Nambu brackets are the tool to describe the classical evolution of systems with  a number of 
constants of motion beyond the minimum required for integrability of the system (see for instance 
\cite{Curtright:2002fd,Curtright:2003fn} and references therein). Geometrically the Nambu structure is related to a 
volume-element Jacobian determinant in the higher dimensional phase space which makes the formalism 
suitable to describe extended objects. This property has been exploited in the search for a formulation of M-theory 
(see for instance \cite{Minic:1999js,Ho:2016hob,Yoneya:2016aja} and references therein). Not less important is the 
relation that Nambu brackets have with several branches of Mathematics such as Algebraic geometry and Group 
theory (see for instance \cite{Naruki2011S170} and references therein).

With the quantization of the rigid body in the three-dimensional phase space as his aim, Nambu identified the 
different Lie structures for the Quantum generators obtaining as result that the possibilities include the $SO(4)$ 
Lie algebra and the ones related to it via analytic continuation or group contraction \cite{Nambu:1973qe}.  In 
other words, the Lie algebras of the generators are the ones corresponding to the different subgroups of 
$SL(3,\mathbb{R})$. Some years later Holm and Marsden studied and classified the classical $SL(2,\mathbb{R})$ 
gauge symmetries  \cite{Marsden1}. Explicitly, they built new Casimir's using  $SL(2,\mathbb{R})$ combinations of 
the original ones $E$ and $G$, concluding that the Hamiltonian form of these combinations can be written as a 
Lie-Poisson system associated to a Lie algebra structure on $\mathbb{R}^3$ and establishing that the different  
Lie algebras are $SO(3)$, $SO(2,1)$, $ISO(2)$ and Heis$_3$.  Particularly interesting is the $ISO(2)$ case 
where the phase space of the Eulerian top is filled with invariant elliptic cylinders on each of which the dynamics is, 
in elliptic coordinates, the dynamics of a standard simple pendulum. The quantization of these systems, some times 
referred as extended free rigid bodies, has been reviewed recently finding explicit solutions only for the 
symmetric bodies \cite{Minic:2002pd,Iwai2010501}. Interesting for our purposes is that in this classification the 
authors also identified the algebra $ISO(1,1)$ as a possible Lie-Poisson structure.

The aim of this contribution is to revisit the classification of the classical $SL(2,\mathbb{R})$ gauge transformations 
following an analysis similar to the one made in \cite{Marsden1}, motivated by several reasons. Firstly, the 
analysis in \cite{Marsden1} is not completely explicit, in fact authors focused mainly on the $ISO(2)$ case. 
Secondly, the analysis makes use of the three principal moments of inertia $I_ i$ $(i=1,2,3)$ as independent 
parameters, however it was recognized long ago, that separability of the Laplace operator for the general 
asymmetric Eulerian top can be achieved only if the analysis is made in spheroconal coordinates where one 
introduces related dimensionless inertia parameters $e_i$  ($i=1,2,3$)
\cite{Ittmann,LukacSmoro,Pinna,Pina1999159}. Thirdly, in analogy to the Nambu's result about the 
$SL(3,\mathbb{R})$ canonical transformations where he found that the inequivalent Lie-Poisson structures are the 
$SO(4)$ Lie algebra and all the ones related to it via analytical continuation or group contraction, it is natural to 
expect that the $SL(2,\mathbb{R})$ gauge symmetry allows as a Lie-Poisson structure all the algebras related to 
$SO(3)$ via analytical continuation and group contractions. Indeed this is the conclusion in \cite{Iwai2010501} 
whereas the case $ISO(1,1)$ is not reported in \cite{Marsden1}. As a result of our analysis we have identified this 
algebra as a possible Lie-Poisson structure for an asymmetric extended rigid body and we show that this algebra is 
not possible for a symmetric body. As a by-product, we include the explicit form of the generators for the different 
possibilities of Lie algebras, this is something that for the best of our knowledge is missing in the literature, and in 
this contribution we fill in this gap. We include in our discussion the Bianchi nomenclature for the different Lie 
algebras \cite{Bianchi2001}. Finally it would be very interesting to find the relation between the Nambu dynamics 
and some recent results in the area of atomic physics and asymmetric rotational molecular spectrums 
\cite{Pina,LeyKoo,MendezFragoso2015115,ley2015rotations}. This issue will be addressed elsewhere.

Relevant for our discussion is the geometrical method introduced by Poinsot for visualizing the motion of the 
endpoint of the angular momentum vector $\vec{L}$, which for the Eulerian top takes place in the intersection 
of the ellipsoid of energy $E$ and the sphere of angular momentum $G$. This construction is automatically 
incorporated in the Nambu formulation, because  $\dot{E}=\dot{G}=0$, and therefore both $E$ and $G$ 
are constants of the motion implying that the orbit of the system in phase space is determined as the intersection 
of the two surfaces  $E=$ const. and $G=$ const. Analytically the parameterization of these intersections is given in 
terms of Jacobi elliptic functions whose modulus depend on the values of the three inertia parameter $e_i$ 
($i=1,2,3$) only (see for instance \cite{Pina1999159} and references therein). In fact, working in these coordinates 
is the only known way that allows to quantize the asymmetric free rigid body
\cite{Ittmann,Reiche,King,spence,LukacSmoro}. 
In these coordinates the geometry of the energy ellipsoid  $E$ is replaced by a Casimir function $h$ with the 
geometry of an elliptic hyperboloid that can be either of one or two sheets depending of the numerical value of the 
quotient $2E/G$ or it can also have the geometry of an elliptic cone  or a hyperbolic cylinder in the proper limit 
situations.  This geometrical richness of the surfaces $h$ gives more geometrical possibilities when 
the Poinsot construction is performed. In this regard our analysis is close to the one performed for the Maxwell-Bloch 
system \cite{David1992}.  

Our exposition is self-contained as possible. In section \ref{RigidBody} we introduce the rigid body system and in 
section \ref{EulerDimensionless} we discuss the issue of the Casimir functions. Section \ref{GeomCas} 
exposes the classification of the different geometries of the Casimir functions and section \ref{DimEuler}
deals with the solutions of the Euler equations. In section \ref{Gauge} we discuss the $SL(2,\mathbb{R})$ gauge 
symmetry of the Eulerian top and in \ref{Classification} we give in detail all the different inequivalent Lie algebras 
for which the Euler equations are the equations of motion. We give our conclusions in \ref{conclusions}.


\section{The rigid body} \label{RigidBody}
 
In a body-fix reference frame the torque free rigid body motion is governed by the Euler equations  
\begin{equation}\label{EulerL}
\frac{d \vec{L}}{dt} = \vec{L} \times I^{-1}  \vec{L},
\end{equation}
where $\vec{L}$ is the vector of angular momentum and $I$ the moment of inertia tensor. If  the body-fixed frame is 
oriented to coincide with the principal axes of inertia, the matrix $I$ is diagonal. Without losing generality in the 
following discussion we consider that the principal moments of inertia satisfy the inequality
\begin{equation}\label{OrderI}
I_1 < I_2 < I_3.
\end{equation}
When written in a basis, the $L_i$  components of the angular momentum are the generators of a  $SO(3)$ Lie 
algebra
\begin{equation}\label{LieAlgebra}
[L_i,L_j]=\varepsilon_{ijk} L_k.
\end{equation}
The problem has two integrals of motion, the kinetic energy $E$ and the square of the angular momentum $L^2$, 
which are given in terms of the moments of inertia and the components of $\vec{L}$ as
\begin{equation}
E = \frac{L_1^2}{2I_1}+  \frac{L_2^2}{2I_2}+ \frac{L_3^2}{2I_3}, \label{ConE}
\end{equation}
and
\begin{equation}
L^2 = L_1^2+L_2^2+L_3^2,\label{ConL}
\end{equation}
respectively. In a Euclidean space whose coordinates are the components of the angular momentum, the equation 
for the energy represents an ellipsoid with semiaxes $\sqrt{2EI_1}$, $\sqrt{2EI_2}$ and $\sqrt{2EI_3}$, whereas the 
equation for the conservation of the modulus of the angular momentum represents an sphere of radius $L$. 
Because $2EI_1< L^2 < 2EI_3$, the radius of the sphere has a value between the minimum and maximum values 
of the semiaxes of the ellipsoid. The dynamical  problem is usually solved using the Poinsot construction. When the 
vector $\vec{L}$ moves relative to the axes of inertia of the top, it  lies along the curve of intersection of the surfaces 
$E=$constant and $L^2=$constant. It is clear that the solutions of the Euler equations (\ref{EulerL})  will depend of 
five different parameters, the three moments of inertia, the energy and the square of the angular momentum. 
However it has been shown that the problem can be rewritten in such a way that the solutions will depend   
only on two parameters. We will return to this point afterwards.


\section{Casimir functions}\label{EulerDimensionless}

Because the Nambu dynamics in three dimensions requires of two ``Hamiltonians'' or Casimir functions, we start 
our discussion reminding this issue in our problem at hand. For the asymmetric rigid body the three principal 
moments of inertia are all different and the proper coordinates 
system that describes it are the spheroconal ones \cite{Ittmann}. These are the coordinates implied 
naturally by the two quadratic Casimir functions of the system: $L^2$ and 
$e_0\equiv e_1 L_1^2+e_2 L_2^2+e_3L_3^2$, where the dimensionless coefficients $e_i$'s ($i=1,2,3$) are 
restricted by two conditions as we will discuss below. In the degenerated cases where two principal moments of 
inertia are equal (symmetric case) or the three are equal (spherical case), the spherical coordinates are the ones 
that naturally incorporate the canonical Casimir functions of the system: $L^2$ and $L_3$. Notice that in the 
asymmetric case the two Casimir functions are quadratic in $L_i^2$, whereas in the symmetric case one of them is 
linear. This issue was systematically addressed in \cite{Patera} and because our analysis strongly depends on the 
geometry of the Casimir functions, here we summarize the main steps to determine these functions.

\begin{figure}[h!]
\begin{center}
\includegraphics[scale=0.2]{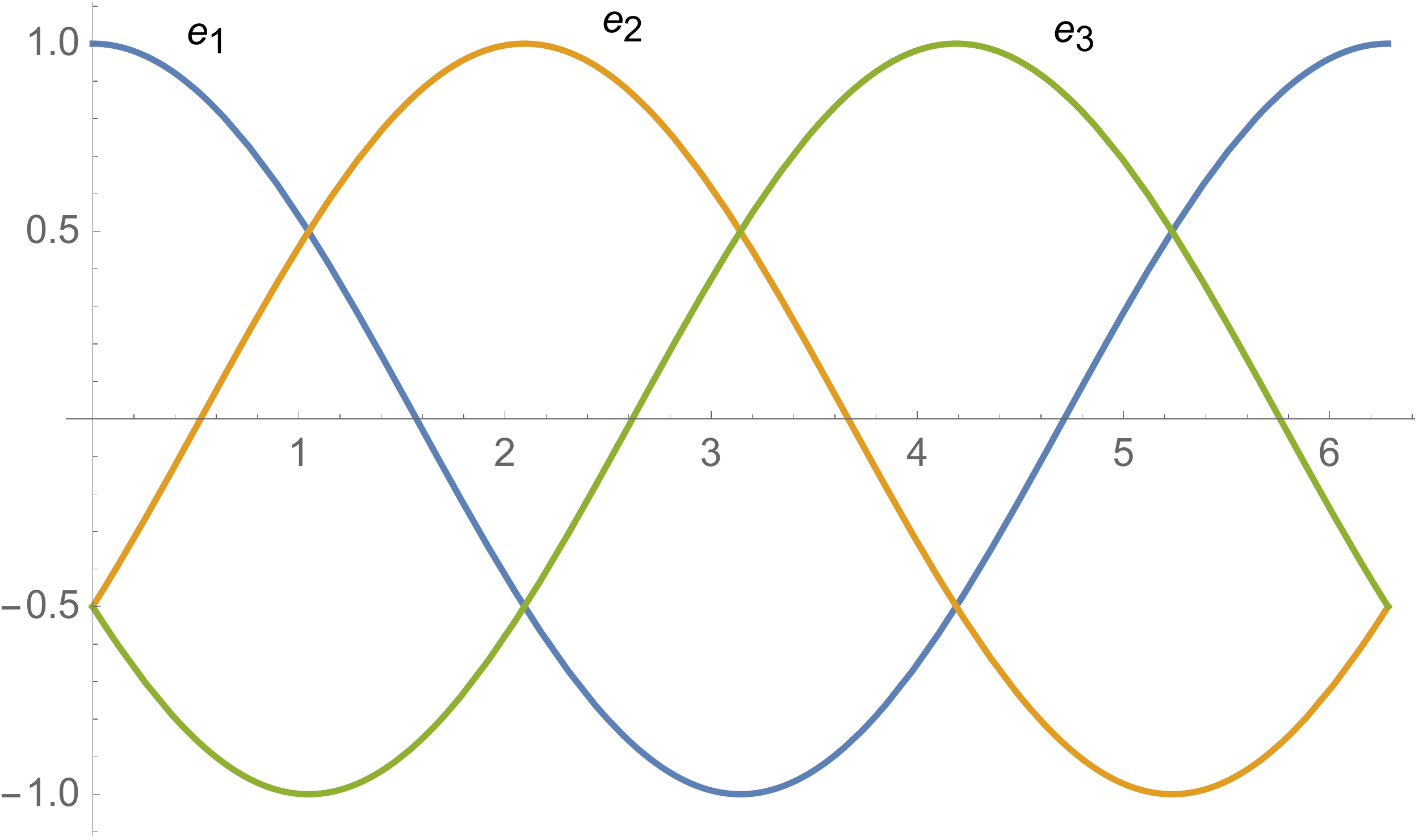}
\end{center}
\caption{Dimensionless inertia parameters $e_1$, $e_2$ and $e_3$ for different values of the asymmetry parameter 
$\kappa\in[0,2\pi]$.}\label{figuraei}
\end{figure}

The starting point is to recognize that any first order vector in the $O(3)$ algebra can be rotated into say $L_3$, and 
therefore  the basis: $L^2$, $L_3$, will constitute a canonical basis. However the asymmetric case is not properly 
described by this basis and in order to find the proper basis, it is necessary to consider symmetric second order 
tensors of the form
\begin{equation}\label{DefSecTen}
N= \sum_{ij} N_{ij} L_i L_j,
\end{equation}
in the enveloping algebra $O(3)$, an example of such a tensor is the kinetic energy $E$ (\ref{ConE}). Here $N$ is a 
real symmetric matrix and working in the representation where it  is diagonal, their eigenvalues are given by  the 
inverse of the principal moments of inertia $N_{ii} \equiv 1/I_i$. The classification depends on the number of 
different eigenvalues which leads to three different cases, two of which are described by  the canonical basis, and 
the remaining one by a basis of two quadratic functions. The cases are the following:
\begin{itemize}
\item  {\it Spherical case:} If $I_1=I_2=I_3$, the tensor $N=L^2/I_1$ is written in terms of the canonical basis. 
\item {\it Symmetric case:} Without losing generality, if  $I_1=I_2\neq I_3$,  the tensor $N=L^2/I_1+L_3^2/(I_3-I_1)$ is 
written in terms of the canonical basis. 
\item {\it Asymmetric case:}  If $I_1 \neq I_2\neq I_3$ the tensor $N=L^2/I_1+(1/I_1-1/I_3) (k_2^2L_2^2+L_3^2)$, and 
the basis is different from the canonical. In fact in addition to $L^2$ the second quadratic tensors can be taken either 
as:  $k_2^2L_2^2+L_3^2$, $L_1^2 + k_1^2 L_2^2$ or  $k_2^2L_1^2-k_1^2L_3^2$. Here 
\begin{equation}\label{Defk}
k_1^2 \equiv \frac{\frac{1}{I_2}-\frac{1}{I_{3}}}{\frac{1}{I_1}-\frac{1}{I_{3}}}
\hspace{0.3cm}  \mbox{and} \hspace{0.3cm}
k_2^2 \equiv \frac{ \frac{1}{I_{1}} - \frac{1}{I_{2}}}{\frac{1}{I_{1}}-\frac{1}{I_{3}}},
\end{equation}
which takes values in the rank $0< k_1^2<1$, $0< k_2^2<1$ and satisfy $k_1^2+k_2^2=1$.
\end{itemize}
In the asymmetric case it is clear that only one quadratic function besides $L^2$ is needed to complete the basis 
because for instance $L_3^2+k_2^2L_2^2=L^2-(L_1^2+k_1^2L_2^2)$. We emphasizes  that the second tensor 
depends of one real parameter only, for instance $k_1^2$. Following  \cite{LukacSmoro,Pinna} we chose 
to rewrite the second quadratic tensor by expressing the matrix $N$ in terms of its irreducible representations: its 
trace and a traceless symmetric matrix, obtaining
\begin{equation}
N = \left( N_{11} - \frac13 \mbox{Tr}\, N \right) L_1^2 +\left( N_{22} - \frac13 \mbox{Tr}\, N \right) L_2^2 + 
\left( N_{33} - \frac13 \mbox{Tr} \, N \right)  L_3^2 +  \frac13 \mbox{Tr} N \,
L^2 . \label{traza1}
\end{equation}
It is possible to replace the set of 3 principal moments of inertia $I_{i}=1/N_{ii}$ by 3 equivalent {\it inertia parameters} 
($\bar{e}_1,\, \bar{e}_2,\, \bar{e}_3$), which have dimension of the inverse of a moment of inertia. The main 
difference between these two sets is that the $\bar{e}_i$'s can be positive, negative or even zero, whereas the $I_i$'s 
are always positive. Writing the diagonal components of the traceless symmetric part of $N$ as \cite{Pina1999159}
\begin{equation}\label{Defes}
N_{11} - \frac13 \mbox{Tr}N \equiv \bar{e}_1, \hspace{0.5cm}
N_{22} - \frac13 \mbox{Tr}N \equiv  \bar{e}_2, \hspace{0.5cm}
N_{33} - \frac13 \mbox{Tr}N \equiv  \bar{e}_3,
\end{equation}
the traceless condition provides a constraint for the 3 parameters $\bar{e}_i$'s
\begin{equation}\label{Cond1e}
\bar{e}_1+\bar{e}_2+\bar{e}_3=0.
\end{equation}
The second constraint on the inertia parameters is obtained by computing the square of the traceless matrix and then
computing its trace, which leads to the condition
\begin{equation}\label{CasiCond2}
\frac13\left[  \left( N_{11}-N_{22} \right)^2 +\left( N_{22}-N_{33} \right)^2  +
\left( N_{33}-N_{11} \right)^2 \right]  =  (\bar{e}_1^2+\bar{e}_2^2+\bar{e}_3^2).
\end{equation}
These two conditions on the inertia parameters are respectively the expressions for both the first and second 
order symmetric invariant functions of the roots of a third order polynomial equation.  The invariance is respect to 
the Galois group which for a third order polynomial is the symmetric group $S_3$ or the permutation group 
$P_3$. Using the standard notation of the Weierstrass elliptic functions, a third order polynomial equation can 
always be cast after a Tschirnhaus transformation to the form 
$4y^3-g_2y-g_3=4(y-\bar{e}_1)(y-\bar{e}_2)(y-\bar{e}_3)=0$, in fact the transformation (\ref{Defes}) is 
a Tschirnhaus transformation applied to the cubic characteristic polynomial associated to the matrix $N_{ij}$ of 
the equation (\ref{DefSecTen}) (see for instance \cite{Gilmore:2008zz}). The three symmetric invariant functions of 
the roots are equation (\ref{Cond1e}), 
$g_2=-4(\bar{e}_1\bar{e}_2+\bar{e}_2\bar{e}_3+\bar{e}_3\bar{e}_1)=2(\bar{e}_1^2+\bar{e}_2^2+\bar{e}_3^2)$ 
and $g_3=4\bar{e}_1\bar{e}_2\bar{e}_3$. The cubic polynomial has three real roots only if $g_2>0$ which is clearly 
satisfied by the change (\ref{Defes}) because from (\ref{CasiCond2})
\begin{equation}\label{DefA}
g_2 \equiv  \frac{2}{3} \left[  \left( \frac{1}{I_1}- \frac{1}{I_2} \right)^2 + \left( \frac{1}{I_2}- \frac{1}{I_3} \right)^2   +
\left( \frac{1}{I_3}- \frac{1}{I_1} \right)^2  \right] >0 .
\end{equation}
The third order symmetric invariant function $g_3$ can have either sign or it can be zero. It is well known that the 
roots $\bar{e}_i$'s can be written in terms of the invariants of the theory. Because the invariant of first order is zero, 
the solutions are given in terms of the invariants $g_2$ and $g_3$ in the form (see for instance \cite{Neumark})
\begin{equation}
\bar{e}_1= -\sqrt{\frac{g_2}{3}}\cos\left(\frac{\varphi}{3} \right) \,, \qquad 
\bar{e}_2= \sqrt{\frac{g_2}{3}}\cos\left(\frac{\varphi+\pi}{3} \right) \,,  \qquad  
\bar{e}_3= \sqrt{\frac{g_2}{3}}\cos\left(\frac{\varphi-\pi}{3} \right)\, ,
\end{equation}
where the angular variable $\varphi \equiv \arccos (-(3/g_2)^{3/2} g_3)$. This form of the solutions is related to the 
ones in \cite{Pina1999159}  by introducing the  {\it asymmetry parameter} $\kappa=\pi+\varphi/3$. Even more, 
we can introduce a related set of {\it dimensionless inertia parameters} $e_i$'s trough the relation 
$\bar{e}_i \equiv \sqrt{g_2/3} \, e_i$, namely
\begin{equation}\label{eidek}
e_1= \cos(\kappa) \,, \qquad e_2= \cos(\kappa-2\pi/3)\,,  \qquad  e_3= \cos(\kappa+2\pi/3)\, ,
\end{equation}
which satisfy the conditions
\begin{equation}\label{CondDimless}
e_1+e_2+e_3=0, \hspace{0.5cm} e_1^2+e_2^2+e_3^2=3/2 \hspace{0.3cm} \mbox{and} \hspace{0.3cm} 
e_1e_2e_3=(\cos 3 \kappa) /4 .
\end{equation}
In this way we obtain finally that any second order tensor $N$ can be expressed as a combination of the two 
quadratic Casimir functions $L^2$ and $e_1 L_1^2 + e_2 L_2^2 + e_3 L_3^2$ in the way
\begin{equation}\label{QaudTensor}
N =  \frac13 \mbox{Tr} N \, L^2 + \sqrt{\frac{g_2}{3}} (e_1 L_1^2 + e_2 L_2^2 + e_3 L_3^2).
\end{equation}
From the geometrical point of view, in the three dimensional space of dimensionless inertia parameters $e_i$'s, for 
the general asymmetric case the first  condition in (\ref{CondDimless}) defines a plane that cross the origin, whereas 
the second one represents a sphere of square radius equal to $3/2$. The intersection of these two surfaces is a 
circle which is parameterized by the angular asymmetry parameter $\kappa$. The role of the invariant $g_2$ is to 
change the size of the sphere and therefore of the intersecting circle whereas the role of the invariant $g_3$ is to 
parameterize the angular variable $\kappa$ that describes the intersecting circle. It is clear that given a set of values 
for the three principal moments of inertia $I_i$, the values of Tr$N$, $g_2$ and $g_3$, and therefore of the angular 
variable $\kappa$ are completely determined. 
Regarding the asymmetry parameter notice that  when it takes values in the interval $\kappa\in[0,\pi/3]$, the inertia 
parameters have the order
\begin{equation}\label{esOrdering}
e_3\leq e_2\leq e_1,
\end{equation}
with $e_3\neq e_1$ (equivalent to condition (\ref{OrderI})). In general as expected, for $\kappa\in[0,2\pi]$, we obtain 
all the six possible orders for the three inertia parameters  $e_i$'s, due to the fact that the inertia parameters are 
given in terms of invariants under the Galois symmetric group $S_3$ (see Figure \ref{figuraei}) and the group is of 
order six. In particular, when $\kappa=n\pi/3$, $n=1,\dots\,, 6$, the rigid body is symmetric (two inertia parameters 
equal), and when $\kappa=(2n-1)\pi/6$, $n=1,\dots\,, 6$ the rigid body is the most asymmetric one (one inertia 
parameter null). For any value of $\kappa$, at least one of the inertia parameters is positive, another is negative, 
and the third one may be either positive, null or negative.

In terms of the $e_i$'s the parameters $k_1^2$ and $k_2^2$ defined in (\ref{Defk}) are 
rewritten respectively as 
\begin{equation}\label{defke}
k_1^2=\frac{e_2-e_3}{e_1-e_3} ,
 \hspace{0.5cm} \mbox{and}  \hspace{0.5cm}
k_2^2=\frac{e_1-e_2}{e_1-e_3}.
\end{equation}
A consequence of a change in the order of the inertia parameters is a change in the expression for these quotients. 
For instance, if the asymmetry parameter takes values in the interval $\kappa\in(\pi/3, 2\pi/3)$ the new parameters 
ordering is $e_3 < e_1 <e_2$ and the parameters analogous to (\ref{defke}) are obtained from them by 
simply interchange $e_1 \leftrightarrow e_2$, obtaining $(e_1-e_3)/(e_2-e_3)=1/k_1^2$ and 
$(e_2-e_1)/(e_2-e_3)=-k_2^2/k_1^2$. In the following table we present the six cases.
\begin{table}[h!] 
\renewcommand*{\arraystretch}{1.5} 
\begin{tabular}{| c | c | c | c |} 
\hline
\hspace{.2cm}Asymmetry parameter interval  \hspace{.2cm} & \hspace{.2cm} Order of the inertia parameters 
\hspace{.2cm}& $\hspace{.2cm}  \hspace{.2cm}$ & $\hspace{.2cm}  \hspace{.2cm}$ \\
\hline 
\hline
$\kappa \in (0,\pi/3 )$ & $e_3<e_2<e_1$ &\hspace{.2cm}  $k_1^2$ \hspace{.2cm} &
\hspace{.2cm} $k_2^2$  \hspace{.2cm}\\ 
\hline 
$\kappa \in (\pi/3,2\pi/3)$ & $e_3<e_1<e_2$ &   $\frac{1}{k_1^2} $& 
$-\frac{k_2^2}{k_1^2}$ \\
\hline 
$\kappa \in (2\pi/3,\pi )$ &  $e_1<e_3<e_2$ & $\frac{1}{k_2^2} $& 
$-\frac{k_1^2}{k_2^2}$  \\
\hline 
$\kappa \in (\pi,4\pi/3 )$ & $e_1<e_2<e_3$ & $k_2^2 $& 
$k_1^2 $ \\
\hline 
$\kappa \in (4\pi/3,5\pi/3 )$ & $e_2<e_1<e_3$ & $-\frac{k_2^2}{k_1^2}$ & 
$\frac{1}{k_1^2} $ \\ 
\hline 
$\kappa \in (5\pi/3,2\pi)$ & $e_2<e_3<e_1$ &$-\frac{k_1^2}{k_2^2} $& 
$\frac{1}{k_2^2} $ \\ 
\hline
\end{tabular}
\caption{List of the six different ordering of the inertia parameters for different intervals of the asymmetry parameter 
$\kappa$. We include the change in the quotients (\ref{defke}). Notice that in every case the analogous restriction to
$k_1^2+k_2^2=1$ takes place.}\label{orderings}
\end{table}


\section{The Geometry of the Casimir functions}\label{GeomCas}

Notice that the quotient $E/L^2$ has dimension of  the inverse of moment of inertia, therefore we can define the 
dimensionless parameter  $e_0$ associated  with the energy and the square of the angular momentum in a 
similar form to (\ref{Defes})  as 
\begin{equation}\label{Defe0}
\frac{2E}{L^2} = \frac13 \mbox{Tr} N+\sqrt{\frac{g_2}{3}}e_0.
\end{equation}
We call to $e_0$ the energy parameter for short. Introducing the unitary vector $\vec{u}$ in the direction of the 
angular momentum $\vec{L}$: $\vec{u} \equiv \vec{L}/ L$ we can rewrite the angular momentum sphere (\ref{ConL}) 
as
\begin{equation}
u_1^2+u_2^2+u_3^2=1.                                                        \label{Con2L}
\end{equation}
Taking $N=2E$ in equation (\ref{QaudTensor}) and combining it with (\ref{Defe0}), we obtain the dimensionless 
form of the another second order Casimir function
\begin{equation}
 e_1 u_1^2 + e_2 u_2^2+ e_3u_3^2=e_0.
\label{Con2E}
\end{equation}
In the dimensionless angular momentum space ($u_1,u_2,u_3$), the surface (\ref{Con2L})  is a unitary sphere, 
whereas the geometry of the Casimir  function (\ref{Con2E}) depends on the number of positive and negative 
coefficients in the equation. In the following we give a classification of the different possible geometries for this 
surface. We emphasize that this surface does not correspond to the energy surface (\ref{ConE}) but is related to it 
through the transformation (\ref{Defe0}). Nevertheless abusing of the language we call to $e_0$ the dimensionless 
energy parameter and to the Casimir function (\ref{Con2E}) the energy surface. A close classification 
to the one we present here was discussed in the context of quantization using the principle of correspondence in
\cite{King}.

\subsubsection{Asymmetric rigid bodies}

In the following without loosing generality we will assume that the inertia parameters satisfy 
inequalities (\ref{esOrdering}).  The condition for intersection between the sphere of angular momentum and the 
energy surface becomes in these parameters 
\begin{equation}\label{e3e0e1}
e_3< e_0 < e_1 \, . 
\end{equation} 
The geometry of the energy surface depends on the value of the energy parameter $e_0$ respect to 
the inertia parameter $e_2$. There are two different cases and they correspond to:
\begin{itemize}
\item{Case $e_2 < e_0 < e_1$ }
\end{itemize}

Generically the surface (\ref{Con2E}) is a hyperboloid instead of the original ellipsoid of energy (\ref{ConE}), because 
at least one inertia parameter is positive ($e_1$ for the ordering (\ref{OrderI})) and another is negative ($e_3$). 
As can be seen in figure \ref{figuraei}, the parameter $e_2$ can be either, positive, zero or negative. Therefore 
because $e_2<e_0$ the energy parameter can also be  either positive, zero or negative. In table \ref{table1} we 
give all different possible geometries of the Casimir function (\ref{Con2E}).

\begin{table}[htb] 
\begin{tabular}{|c | c | c | c |} 
\hline
$e_2$ & $e_0$ & ``Energy" surface equation & ``Energy" surface geometry\\
\hline 
\hline
$ < 0$ & $<0$ & $-\frac{e_1}{|e_0|}u_1^2+\frac{|e_2|}{|e_0|}u_2^2 +\frac{|e_3|}{|e_0|}u_3^2=1$& 
elliptic hyperboloid of one sheet  around $u_1$ \\ 
$<0$ &  $=0$ & $ e_1u_1^2= |e_2| u_2^2 + |e_3| u_3^2$  &elliptic cone around $u_1$ \\
$<0$ & $>0$ &  $\frac{e_1}{e_0}u_1^2-\frac{|e_2|}{e_0}u_2^2 - \frac{|e_3|}{e_0}u_3^2=1$ 
& elliptic hyperboloid of two sheets around $u_1$\\
$=0$ & $>0$ &   $\frac{e_1}{e_0}(u_1^2 - u_3^2)=1$ & hyperbolic cylinder \\
$>0$ & $>0$ &  $\frac{e_1}{e_0}u_1^2+\frac{e_2}{e_0}u_2^2 - \frac{|e_3|}{e_0}u_3^2=1$ & 
elliptic hyperboloid of one sheet around $u_3$\\ 
\hline
\end{tabular}
\caption{List of the different geometries the Casimir function (\ref{Con2E}) can describe, in the case where the 
dimensionless energy parameter $e_0$ take values in the interval $e_2 < e_0 < e_1$.}\label{table1}
\end{table}

\begin{itemize}
\item{Case $e_3 < e_0 < e_2$ }
\end{itemize}

In this case the energy surfaces are very similar to the previous case. The main difference is the orientation 
of the surfaces. In table \ref{table2} we list the different  geometries of the energy surface.

\begin{table}[htb] 
\begin{tabular}{|c | c | c | c |} 
\hline
$e_0$ & $e_2$ & ``Energy" surface equation & ``Energy" surface geometry\\
\hline 
\hline
 $ < 0$ & $<0$ & $-\frac{e_1}{|e_0|}u_1^2+\frac{|e_2|}{|e_0|}u_2^2 +\frac{|e_3|}{|e_0|}u_3^2=1$& 
 elliptic hyperboloid of one sheet  around $u_1$ \\ 
 $<0$ &  $=0$ &    $-\frac{e_1}{|e_0|}(u_1^2 - u_3^2)=1$ & hyperbolic cylinder \\
 $<0$ & $>0$ &  $-\frac{e_1}{|e_0|}u_1^2-\frac{e_2}{|e_0|}u_2^2 + \frac{|e_3|}{|e_0|}u_3^2=1$ 
 & elliptic hyperboloid of two sheets around $u_3$\\
 $=0$ & $>0$ &   $ e_1 u_1^2 + e_2 u_2^2=|e_3|u^2_3$ & elliptic cone around $u_3$\\
 $>0$ & $>0$ &  $\frac{e_1}{e_0}u_1^2+\frac{e_2}{e_0}u_2^2 - \frac{|e_3|}{e_0}u_3^2=1$ & 
 elliptic hyperboloid of one sheet around $u_3$ \\ 
\hline
\end{tabular}
\caption{List of the different geometries the Casimir function (\ref{Con2E}) can describe, in the case where the 
dimensionless energy parameter $e_0$ take values in the interval $e_3 < e_0 < e_2$.}\label{table2}
\end{table}


\subsubsection{Symmetric rigid bodies}

Important throughout our exposition are the special cases when the rigid body has cylindric symmetry. These cases  
can be obtained as limiting situations of the general asymmetric rigid body taking two inertia parameters equal. 
There are two of such limits in the parameters ordering (\ref{esOrdering}) we are discussing here. The 
so called {\it prolate} limit $e_3=e_2$ and the {\it oblate} limit $e_2=e_1$. 

\begin{itemize}
\item{Prolate case, $e_3=e_2<0<e_1\, \, ( \kappa=0$)}: The smallest principal momentum of inertia is $I_1$. We can 
take the  prolate limit in 3 different situations. The geometries of the Casimir function (\ref{Con2E}) are given in table 
\ref{table1Sym}.

\begin{table}[h!]  
\begin{tabular}{| c | c | c |} 
\hline
$e_0$ & ``Energy" surface equation & ``Energy" surface geometry \\
\hline 
\hline
$<0$ & $-\frac{e_1}{|e_0|}u_1^2+\frac{|e_3|}{|e_0|}( u_2^2 + u_3^2)=1$& 
hyperboloid of revolution of one sheet  around $u_1$ \\ 
 $=0$ & $ e_1u_1^2= |e_3| (u_2^2 + u_3^2)$  & circular cone around $u_1$ \\
 $>0$ &  $\frac{e_1}{e_0} u_1^2-\frac{|e_3|}{e_0}(u_2^2 + u_3^2)=1$ 
 & hyperboloid of revolution of two sheets around $u_1$\\
\hline
\end{tabular}
\caption{List of the different geometries that the Casimir surface (\ref{Con2E}) can take in the symmetric prolate case.}
\label{table1Sym}
\end{table}
\end{itemize}

\begin{itemize}
\item{Oblate case, $e_3<0<e_2=e_1\, \, (\kappa=\pi/3$)}: The largest principal momentum of inertia momentum 
$I_3$. The oblate limit also occurs in 3 different situations. The geometries of the Casimir function (\ref{Con2E}) are 
given in the table \ref{table2Sym}.
\begin{table}[h!]  
\begin{tabular}{| c | c | c |} 
\hline
$e_0$ & ``Energy" surface equation & ``Energy" surface geometry \\
\hline 
\hline
$<0$ & $-\frac{e_1}{|e_0|}(u_1^2 + u_2^2) + \frac{|e_3|}{|e_0|}u_3^2=1$ 
& hyperboloid of revolution of two sheets around $u_3$ \\ 
$=0$ & $ e_1 (u_1^2 + u_2^2)=|e_3|u^2_3$ & circular cone around $u_3$ \\
$>0$ &  $\frac{e_1}{e_0}(u_1^2+ u_2^2) - \frac{|e_3|}{e_0}u_3^2=1$ & 
hyperboloid of revolution of one sheet around $u_3$ \\
\hline
\end{tabular}
\caption{List of the different geometries that the Casimir surface (\ref{Con2E}) can take in the symmetric oblate case.}
\label{table2Sym}
\end{table}
\end{itemize}


\section{Dimensionless Euler equations and its solutions}\label{DimEuler}

It is straightforward to write down the Euler equations  (\ref{EulerL}) in terms of both the dimensionless inertia 
parameters $e_i$'s and the dimensionless vector of angular momentum $\vec{u}$. In components form we have
\begin{equation}
\frac{d u_i}{dt}  = \frac12  \sqrt{\frac{g_2}{3}}  L \, \varepsilon_{ijk} (e_k-e_j) u_j u_k .
\end{equation}
It is possible to absorb the factor $ \sqrt{\frac{g_2}{3}}  L$, by defining a dimensionless time parameter: 
$x \equiv t  \sqrt{\frac{g_2}{3}}  L$, obtaining finally
\begin{equation}\label{euleradim}
\dot{u}_i =\frac12  \varepsilon_{ijk} (e_k-e_j) u_j u_k \,,
\end{equation}
where ($\cdot$) denotes derivative with respect to the dimensionless time $x$.
The solutions to these equations in the general asymmetric case are given in terms of elliptic functions. From 
the classical mechanics point of view there are at least two different but equivalent ways 
to solve them. One method of solution consists in  applying the Poinsot construction, {\it i.e.} the solutions are 
obtained as the intersection of the energy and angular momentum surfaces.
In \cite{Landau} the solutions are obtained for the Casimir functions (\ref{ConE})-(\ref{ConL}) whereas in 
\cite{Pinna} the solutions are obtained for the Casimir functions
(\ref{Con2L})-(\ref{Con2E}). The second form of solving the equations is known as the separation method  
(see for instance \cite{Pinna2,PINAGARZA2008} and references therein) which is the relevant method when one 
deals with the quantization of the asymmetric rigid body. Here we are not going into details, but for purposes of 
completeness in our exposition here we give the explicit solution obtained by both methods, details can be found for 
instance in \cite{Pinna,Pinna2}.
\begin{itemize}
\item[{\bf I.}] Case $e_3<e_2<e_0<e_1 $ \, ({\it i.e.} $1/I_2<2E/L^2$).
\end{itemize}
When the energy parameter $e_0$ is between the inertia parameters $e_2$ and $e_1$, the solutions are given by 
\begin{equation}\label{Solutions1}
u_1(\tau)=  \mbox{sn}(\tau ' ,m_c) \, \mbox{dn}(\tau,m), \hspace{0.5cm}
u_2(\tau)= \mbox{dn}(\tau ' ,m_c)  \, \mbox{sn}(\tau,m), \hspace{0.5cm}
u_3 (\tau)= \mbox{cn}(\tau ' ,m_c) \,  \mbox{cn}(\tau,m),
\end{equation}
where $\tau$ is a dimensionless time parameter defined as
\begin{equation}\label{Deftau}
\tau= x \sqrt{(e_1-e_2)(e_0-e_3)}.
\end{equation}
Here the amplitudes of the solutions are written as Jacobi elliptic functions at parameter $\tau ' =$ constant and are 
related to the dimensionless parameters $e_i$ and $e_0$ in the form
\begin{equation}\label{identifications}
\mbox{sn}^2(\tau ' ,m_c) = \frac{e_0-e_3}{e_1-e_3} , \hspace{0.5cm}
\mbox{cn}^2(\tau ' ,m_c) = \frac{e_1-e_0}{e_1-e_3} , \hspace{0.5cm}
\mbox{dn}^2(\tau ' ,m_c)  = \frac{e_1-e_0}{e_1-e_2}.
\end{equation}
The square modulus of the elliptic functions is
\begin{equation}\label{equ}
m^2\equiv \frac{(e_2-e_3)}{(e_0-e_3)}  \frac{(e_1-e_0)}{(e_1-e_2)}  = \frac{(e_1-e_0)}{(e_0-e_3)}\, 
\frac{k_1^2}{k_2^2}\,.
\end{equation}
which takes values in the interval $0<m^2<1$. $m_c$ is its complementary modulus $m_c^2=1-m^2$.

\begin{itemize}
\item[{\bf II.}] Case $e_3<e_0<e_2<e_1$ \, ({\it i.e.}  $ 2E/L^2<1/I_2$).
\end{itemize}
In this case the solutions are given by 
\begin{equation}\label{Solutions2}
u_1=\mbox{cn}\left( \tau ' , i\frac{m_c}{m} \right) \,  \mbox{cn}\left( m\tau,\frac{1}{m} \right), \quad
u_2=\mbox{dn} \left( \tau ' , i\frac{m_c}{m} \right)  \, \mbox{sn} \left( m\tau,\frac{1}{m} \right), \quad
u_3 =\mbox{sn}\left( \tau ' , i\frac{m_c}{m} \right) \, \mbox{dn}\left( m\tau,\frac{1}{m} \right).
\end{equation}
where the square modulus of the elliptic function take values in the interval
$0<1/m^2 < 1$. Here $m^2$ was defined in (\ref{equ}) but due to the fact that $e_3<e_0<e_2<e_1$, 
$m^2>1$. The time parameter  $\tau$ is the one defined in (\ref{Deftau}).

Physical interpretation of the solutions is straightforward, the curves (\ref{Solutions1})-(\ref{Solutions2}) are the 
parameterization of the intersection of the unitary sphere of angular momentum and the corresponding surface 
of the Casimir function (\ref{Con2E}), which for the asymmetric case is given in tables \ref{table1} and \ref{table2}. 
It is important to accentuate that both the dimensionless time parameter $\tau$ and the modulus $m$ in the elliptic 
functions depend explicitly on the inertia parameters $e_i$'s and the energy parameter $e_0$. 
This fact implies for instance that the parameterization of 
the curves on the angular momentum sphere (\ref{Con2L}) depends on the value of $e_0$. However, in principle it  
should be possible to parameterize these curves in terms of elliptic functions whose modulus does not depend 
of this parameter, because the sphere represents the surface of angular momentum only. This parameterization
exists and it is mandatory to use it in the quantization of  
the asymmetric top because in this basis, it is possible to achieve separability of the 
Laplacian operator in the Liouville sense (see for instance \cite{Pinna2} for a discussion of this property). 
For this reason the solution method is called the separation method.

The parameterization of the solutions is obtained by using a first degree transformation of the Jacobi elliptic 
functions  \cite{Akhiezer}.  A detailed derivation of the change of basis is discussed in 
\cite{Pinna2,PINAGARZA2008}, and is given explicitly by
\begin{equation}\label{trans}
\mbox{sn}(\tau_1,m)=\frac{k_2 \, \mbox{sn}(\chi_1,k_1)}{\mbox{dn}(\chi_1,k_1)} ,
\hspace{0.5cm} \mbox{and} \hspace{0.5cm}
m \, \mbox{sn}(\tau_2,m)= - \frac{k_1 \, \mbox{sn}(\chi_2,k_2)}{\mbox{dn}(\chi_2,k_2)}.
\end{equation}
There are similar transformations for the other two Jacobi functions cn and dn. In the new parameterization the 
time parameter $\tau$ (\ref{Deftau}),  is written as a difference of two new time parameters  
$\tau_1$ and $\tau_2$, {\it i.e.}  $\tau=\tau_1-\tau_2$ and the new angular parameters $\chi_1$ and $\chi_2$ 
depend on the dimensionless inertia parameters as
\begin{equation}
\chi_1\equiv\sqrt{e_1-e_3}\,  \phi_1,  \hspace{0.5cm} \mbox{and}  \hspace{0.5cm}
\chi_2\equiv\sqrt{e_1-e_3}\,  \phi_2 .
\end{equation}
The modulus $k_1$ and $k_2$ of the elliptic functions are given by equations (\ref{defke}). Substitution of 
transformations (\ref{trans}) plus the ones for the other Jacobi functions into equations (\ref{Solutions1})
leads to a parameterization of the unitary $S^2$ sphere in terms of spheroconal coordinates that does not depend 
explicitly on $e_0$
\begin{equation}\label{spheroconal}
u_1 = \, \mbox{dn}(\chi_1,k_1)\, \mbox{sn}(\chi_2,k_2), \hspace{0.5cm}
u_2 =  \, \mbox{cn}(\chi_1,k_1)\, \mbox{cn}(\chi_2,k_2), \hspace{0.5cm}
u_3 = \, \mbox{sn}(\chi_1,k_1)\, \mbox{dn}(\chi_2,k_2),
\end{equation}
where the angular parameters are defined in the intervals
\begin{equation}\label{intervals}
-2K_1 \leq \chi_1 \leq 2K_1, \hspace{0.5cm} \mbox{and}  \hspace{0.5cm} -K_2 \leq \chi_2 \leq K_2,
\end{equation}
with $K_1$ and $K_2$ the corresponding quarter periods of the Jacobi elliptic functions \cite{Patera}. 
Because
\begin{equation}
\frac{u_1^2}{\mbox{dn}^2(\chi_1,k_1)} + \frac{u_2^2}{\mbox{cn}^2(\chi_1,k_1)}  = 1, \hspace{0.3cm} \mbox{and} 
\hspace{0.3cm} 
\frac{u_2^2}{\mbox{cn}^2(\chi_2,k_2)} + \frac{u_3^2}{\mbox{dn}^2(\chi_2,k_2)}  = 1,
\end{equation}
the coordinates $\chi_1$ and $\chi_2$ are elliptical coordinates on the sphere with axis along the $u_1$ and 
$u_3$ direction respectively. From here onwards we will work in the parameterization (\ref{spheroconal}).


\section{$SL(2,\mathbb{R})$ symmetries of the Euler equations}\label{Gauge}

In order to make manifest the gauge symmetries of the Euler equations, we notice that its dimensionless  form
\begin{equation}
\dot{\vec{u}}= \vec{u} \times \epsilon \vec{u}
\end{equation}
with $\epsilon$ a diagonal matrix of the form $\epsilon =$diag$(e_1,e_2,e_3)$, can be rewritten as the gradient of 
two scalar functions
\begin{equation}
\dot{\vec{u}}=\nabla l  \times \nabla h,
\end{equation}
where
\begin{eqnarray}
h=\frac12 (e_1u_1^2+ e_2 u_2^2+e_3 u_3^2)- \frac12 e_0 =0 &\Rightarrow&  \epsilon \vec{u}= \nabla h = 
 (e_1u_1,e_2u_2,e_3u_3), \label{Defh}\\
l=\frac12 (u_1^2+ u_2^2+u_3^2) - \frac12 =0 &\Rightarrow& \vec{u}=\nabla l = (u_1,u_2,u_3) \label{Defl}.
\end{eqnarray}
This form of writing the Euler equations makes explicit its invariance under any $SL(2,\mathbb{R})$  transformation 
\cite{Marsden1}.
It is straightforward to check that the transformation
\begin{equation}\label{TransSL2}
 \left( 
\begin{array}{c}
H \\
N
\end{array}
\right) = \left( 
\begin{array}{cc}
a & b \\
c& d
\end{array}
\right)
 \left( 
\begin{array}{c}
h \\
l
\end{array}
\right), 
\end{equation}
with $ad-bc=1$ leads to
\begin{equation}\label{EulerHN}
\dot{\vec{u}}=\nabla N  \times \nabla H.
\end{equation}
The Nambu dynamics established that, given a dynamical system with Hamiltonian $H$, any dynamical 
quantity $Q$ evolves with time according to
\begin{equation}
\dot Q=\{Q,N,H\}=X_HQ,
\end{equation}
where the generator $X_G$ is given by
\begin{equation}
X_G=( \nabla N  \times \nabla G) \cdot \nabla.
\end{equation}
In the spheroconal coordinates (\ref{spheroconal}), the gradient operator has components
\begin{eqnarray} 
\frac{\partial}{\partial u_1}&=& \frac{1}{\tilde{h}^2}\left[ \mbox{dn}(\chi_1,k_1)\mbox{cn}(\chi_2,k_2)
\mbox{dn}(\chi_2,k_2) \frac{\partial}{\partial \chi_1} + k_1^2 \mbox{sn}(\chi_1,k_1)\mbox{cn}(\chi_1,k_1)
\mbox{sn}(\chi_2,k_2) \frac{\partial}{\partial \chi_2} \right], \nonumber \\
\frac{\partial}{\partial u_2}&=&\frac{1}{\tilde{h}^2}\left[ -\mbox{cn}(\chi_1, k_1)\mbox{sn}(\chi_2, k_2)
\mbox{dn}(\chi_2, k_2) \frac{\partial}{\partial \chi_1} + \mbox{sn}(\chi_1, k_1)\mbox{dn}(\chi_1, k_1)
\mbox{cn}(\chi_2, k_2) \frac{\partial}{\partial \chi_2} \right], \label{Killings} \\
\frac{\partial}{\partial u_3}&=&\frac{1}{\tilde{h}^2}\left[ -k^2_2 \mbox{sn}(\chi_1, k_1)\mbox{sn}(\chi_2, k_2)
\mbox{cn}(\chi_2, k_2) \frac{\partial}{\partial \chi_1} + \mbox{cn}(\chi_1, k_1)\mbox{dn}(\chi_1, k_1)
\mbox{dn}(\chi_2, k_2) \frac{\partial}{\partial \chi_2} \right] , \nonumber
\end{eqnarray}
where the function $\tilde{h}$ is defined as
\begin{equation}
\tilde{h}^2\equiv \tilde{h}^2(\chi_1,\chi_2)=1-k_1^2 \mbox{sn}^2(\chi_1,k_1) - k_2^2 \mbox{sn}^2(\chi_2,k_2).
\end{equation}
According with the transformation (\ref{TransSL2}), the generator $X_G$ in components form is expressed as
\begin{equation}
X_G=(ce_1+d)u_1\left( \frac{\partial G}{\partial u_2}\frac{\partial}{ \partial u_3} -  
\frac{\partial G}{\partial u_3}\frac{\partial}{ \partial u_2} \right) + 
(ce_2+d)u_2\left( \frac{\partial G}{\partial u_3}\frac{\partial}{ \partial u_1} -  
\frac{\partial G}{\partial u_1}\frac{\partial}{ \partial u_3} \right) + 
(ce_3+d)u_3\left( \frac{\partial G}{\partial u_1}\frac{\partial}{ \partial u_2} -  
\frac{\partial G}{\partial u_2}\frac{\partial}{ \partial u_1} \right).
\end{equation}
In order to determine the structure of the Lie algebra, we calculate the Lie-Poisson bracket for the Hamiltonian 
vector fields associated with the coordinate functions $u_i$
\begin{eqnarray}
X_{u_1}&=& (ce_3+d) u_3 \partial_2- (ce_2+d) u_2 \partial_3, \nonumber \\
X_{u_2}&=& (ce_1+d) u_1 \partial_3- (ce_3+d) u_3 \partial_1, \label{BasicGen} \\
X_{u_3}&=& (ce_2+d) u_2 \partial_1- (ce_1+d) u_1 \partial_2. \nonumber
\end{eqnarray}
Explicitly the Lie-Poisson brackets  are given by
\begin{equation}\label{GeneralLie}
[X_{u_1},X_{u_2}]=(ce_3+d)X_{u_3}, \hspace{0.5cm} 
[X_{u_2},X_{u_3}]=(ce_1+d)X_{u_1}, \hspace{0.5cm}
[X_{u_3},X_{u_1}]=(ce_2+d)X_{u_2}.
\end{equation}
The classification of the different Lie algebras arise from the $SL(2,\mathbb{R})$ invariant orbits. Before we address 
this point, it is convenient to remind that for the symmetric cases the suitable coordinates are the spherical 
coordinates. We can obtain the corresponding generators from  (\ref{spheroconal}) in the proper limits. For the 
oblate case ($e_1=e_2$) we have, due to the fact that Jacobi elliptic functions go to hyperbolic (circular) functions 
for a value of the modulus $k_1=1$ ($k_2=0$), that the coordinates (\ref{spheroconal}) become
\[
u_1= \sech(\chi_1)\, \sin (\chi_2), \hspace{0.5cm}
u_2= \sech(\chi_1)\, \cos (\chi_2), \hspace{0.5cm}
u_3= \tanh(\chi_1)\, .
\]
Geometrically in this limit the ellipses  $\chi_1$=constant become circles and therefore the 
coordinate $\chi_2$ can be renamed as the standard azimuthal angle $\chi_2\equiv \phi$. The half piece of the 
ellipses $\chi_2$=constant degenerate into the semi-circle intersection of the half plane: $u_2=0$, $u_1>0$, with 
the sphere. Because the hyperbolic function $\sech(\chi_1)$ is always positive unlike the elliptic Jacobi function 
cn$(\chi_1)$, in order to cover the whole sphere, the angle $\phi$ must be 
extended to take values in the interval $[-\pi,\pi)$ instead of the  interval (\ref{intervals}) for $\chi_1$. It is common to 
parametrize the unitary $S^2$ sphere with circular functions instead of hyperbolic ones, thus replacing 
$\sech(\chi_1) \rightarrow \sin(\theta)$ $\Rightarrow$ 
$\tanh(\chi_1) \rightarrow  \cos(\theta)$ with $\theta\in[0,\pi]$, we obtain
\begin{equation}\label{oblate}
u_1= \,  \sin \theta\, \sin \phi, \hspace{0.5cm}
u_2= \,  \sin \theta\, \cos \phi, \hspace{0.5cm}
u_3= \,  \cos \theta\, .
\end{equation}
As a consequence we have from (\ref{Killings}) that the components of the gradient operator in the oblate case are 
\begin{eqnarray}
\partial_1&=&   \sin \phi \cot \theta  \frac{\partial}{\partial \phi}  + \cos \phi \csc \theta \frac{\partial}{\partial \theta}, 
\nonumber \\
\partial_2&=&  \cos \phi \cot \theta \frac{\partial}{\partial \phi}  - \sin \phi \csc \theta \frac{\partial}{\partial \theta}, \\
\partial_3&=& \frac{\partial}{\partial \phi} . \nonumber
\end{eqnarray}
The prolate case ($e_2=e_3$) is obtained in the limit $k_1=0$ and $k_2=1$. Proceeding in a similar way to the 
oblate case, for the prolate rigid body we have that the suitable parameterization of the $S^2$ sphere is
\begin{equation}\label{prolate}
u_1= \,  \cos \theta, \hspace{0.5cm}
u_2= \,  \sin \theta\, \cos \phi, \hspace{0.5cm}
u_3= \,  \sin \theta\, \sin \phi,
\end{equation}
obtaining that the components of the gradient operator  (\ref{Killings}) become
\begin{eqnarray}
\partial_1&=&   \frac{\partial}{\partial \theta}, \nonumber \\
\partial_2&=&  - \cos \theta \cot \phi \frac{\partial}{\partial \theta}  + \sin \theta \csc \phi \frac{\partial}{\partial \phi}, \\
\partial_3&=&  - \sin \theta \cot \theta \frac{\partial}{\partial \theta} + \cos \theta \csc \phi  \frac{\partial}{\partial \phi}. 
\nonumber
\end{eqnarray}


\section{Classification of the $SL(2,\mathbb{R})$ gauge transformations}\label{Classification}

We are now in position to fully classify the $SL(2,\mathbb{R})$ gauge transformations of the Eulerian top. 
The transformations are divided into three general sets where each set belongs to a different fixed form of the 
$SL(2,\mathbb{R})$ matrix. As we will discuss, for some sets there are different cases where the Hamiltonian 
vector fields belong to inequivalent forms of the three dimensional Lie algebras. 
Our analysis follows closely to the one performed in \cite{David1992}.  Along our discussion we 
use both the nomenclature  commonly used in the Lie algebras literature (see for instance \cite{Gilmore:2008zz}) 
and the one introduced by Bianchi in his classification of three dimensional Lie algebra 
\cite{Bianchi2001}. 


\subsection{The Bianchi IX or $SO(3)$ algebra: $c=0$ and $d \neq 0$}

A generic  $SL(2,\mathbb{R})$ group element in this case has the form
\begin{equation}\label{GaugeIX}
g= \left( 
\begin{array}{cc}
a & b \\
0 & 1/a
\end{array}
\right) \hspace{0.3cm} \mbox{with}  
\hspace{0.3cm} 
b \in \mathbb{R}. 
\end{equation} 
Without lossing generality we can assume that $a>0$. These group elements are of the type $AN$ in the Iwasawa 
decomposition of  the special linear group \cite{Sugiura}, and as a generic characteristic they do not change the 
type of group orbit. In this case, we can define $Y_{u_1}\equiv a X_{u_1}$, $Y_{u_2}\equiv  a X_{u_2}$  
and $Y_{u_3}\equiv  a X_{u_3}$  and we obtain the $SO(3)$ Lie algebra
\begin{equation}\label{SO3}
[Y_{u_1},Y_{u_2}]=Y_{u_3}, \hspace{0.5cm} 
[Y_{u_2},Y_{u_3}]=Y_{u_1}, \hspace{0.5cm}
[Y_{u_3},Y_{u_1}]=Y_{u_2}.
\end{equation}
According to equation (\ref{TransSL2}) the transformed $SL(2,\mathbb{R})$ dynamical quantities are
\begin{equation}\label{DefH1}
H=\frac{1}{2} \left[ (ae_1 + b ) u_1^2 + (ae_2+b) u_2^2 + ( ae_3 +b) u_3^2 \right]  - \frac{1}{2}(ae_0+b) =0,
\hspace{0.5cm}  \mbox{and}  \hspace{0.5cm}  
N=\frac{1}{2a} \left[  u_1^2 + u_2^2 +  u_3^2 \right] - \frac{1}{2a}=0 ,
\end{equation}
whereas the Hamiltonian generator is given by
\begin{equation}
Y_G=-(ae_1+b) u_1Y_{u_1} -(ae_2+b) u_2Y_{u_2} -(ae_3+b) u_3Y_{u_3}. 
\end{equation}
Notice that the Casimir $N$  is indeed the same as the Casimir $l$ (equation (\ref{Defl})), and therefore 
geometrically it is represented by an $S^2$ sphere. On the contrary the geometry of the Casimir $H$ and the 
expression of the Hamiltonian generator $Y_G$ depends on the values of the coefficients $ae_i+b$. 
Determining these quantities requires an analysis by cases. The one corresponding to $b=0$ is the simplest case
and describes the evolution of the dimensionless angular momentum of an asymmetric rigid body, because the 
Casimir $H$ is given simply by $h$. According to the Euler equation (\ref{EulerHN}) the physics occurs in the 
intersection of the surfaces $l$ and $h$, and for $b=0$ in the asymmetric situation this occurs in the 10 different 
forms listed in tables \ref{table1} and \ref{table2}, as well as with its 3+3 symmetric limits. 

Exceptional  cases take place if $e_0$ has some of the three values $e_1$, $e_2$ or $e_3$. If $e_0=e_1$ (or 
$e_0=e_3$), the hyperboloid and the sphere are tangent in only two points: $(u_1,u_2,u_3)=(\pm 1,0,0)$ (or  
$(u_1,u_2,u_3)=(0,0,\pm 1)$). But if $e_0=e_2$ the hyperboloid and the sphere are tangent in two big circles 
forming the separatrix, with two opposite unstable points at $(u_1,u_2,u_3)=(0, \pm 1,0,)$.  

\begin{figure}[htb]
\begin{center}
\subfigure[$\, e_0=e_3$]{\includegraphics[scale=0.1]{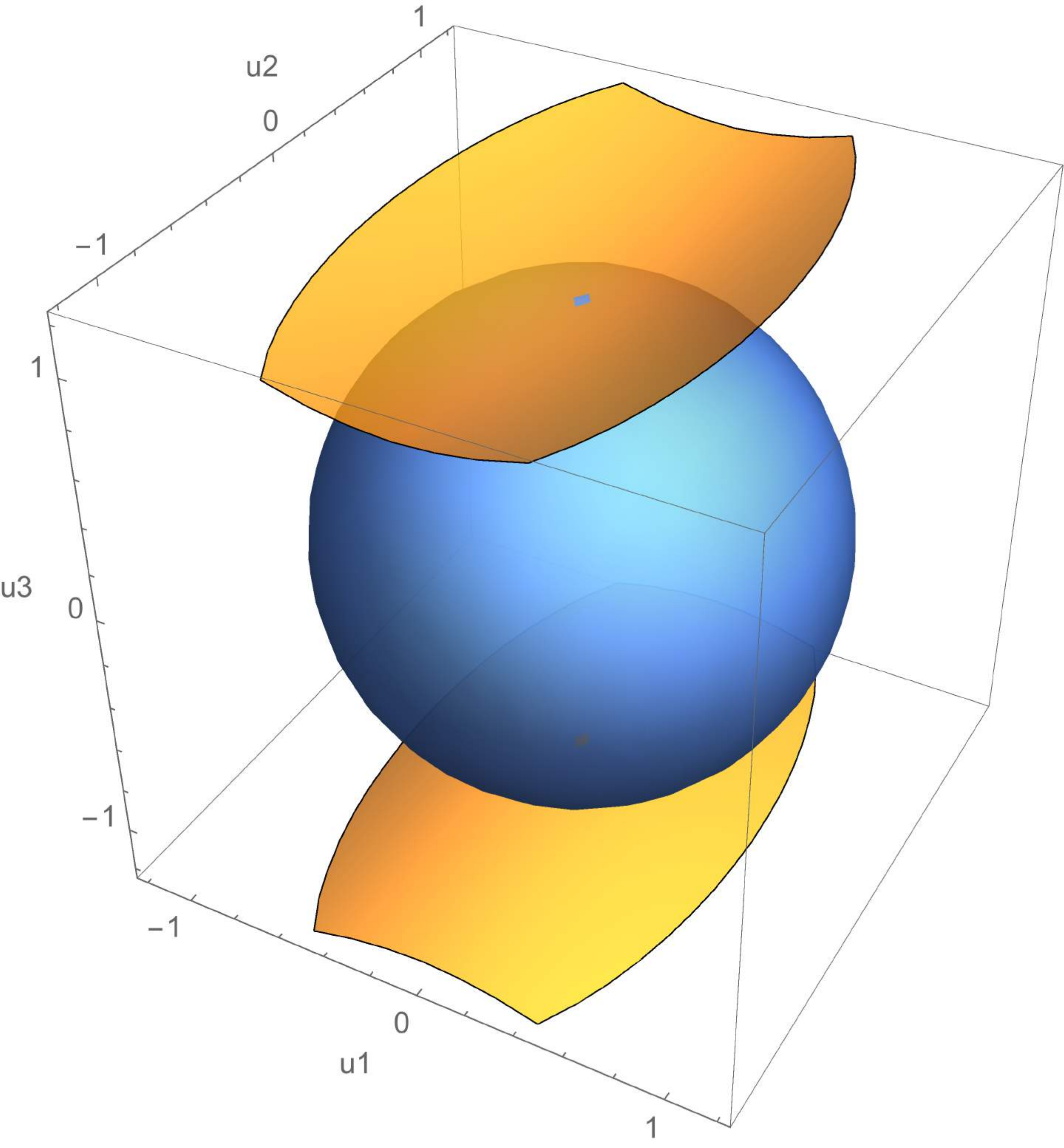}  } \hspace{0.5in}
\subfigure[$\, e_0=e_2$]{\includegraphics[scale=0.1]{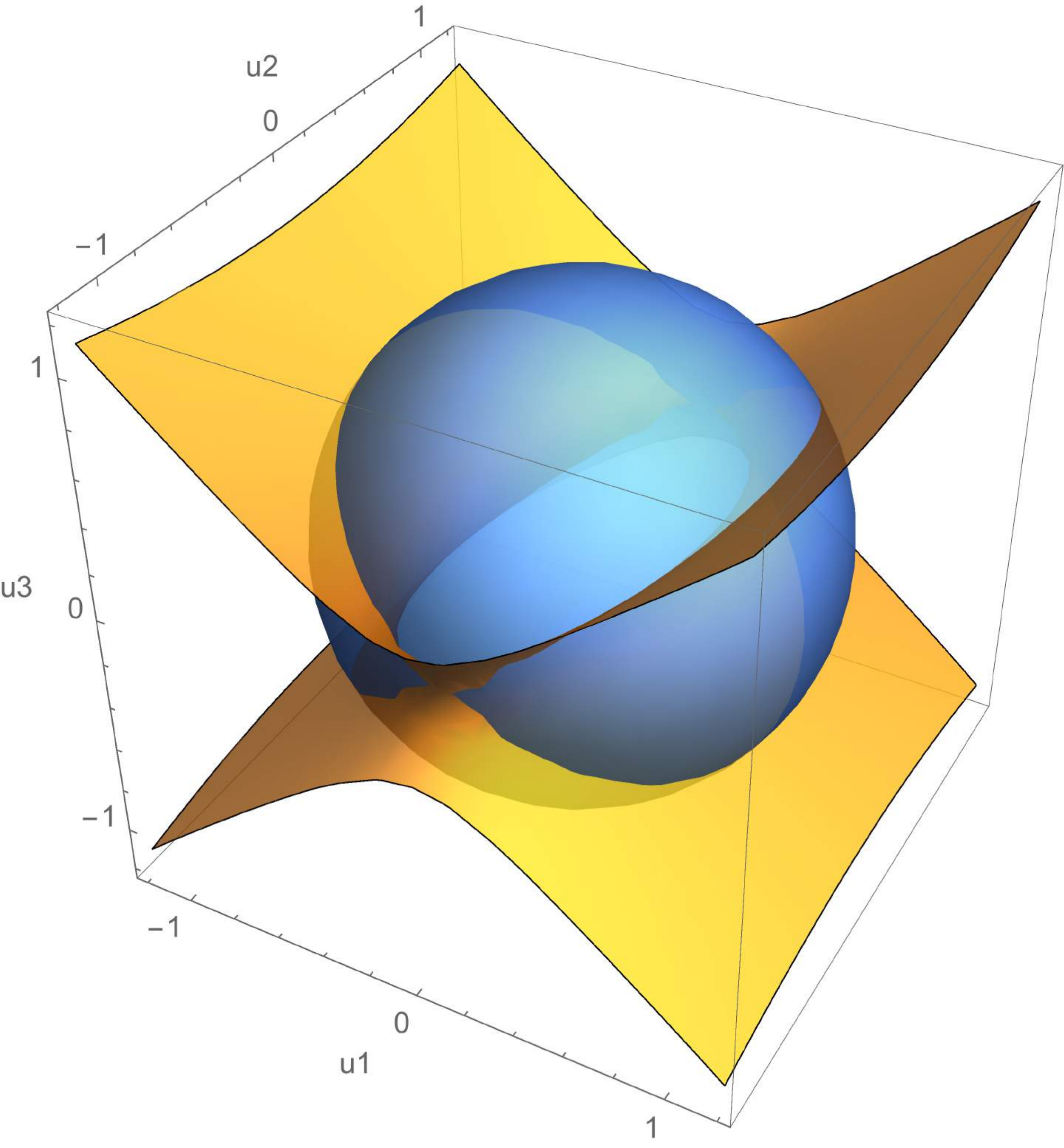}  }\hspace{0.5in} 
\subfigure[$\, e_0=e_1$]{\includegraphics[scale=0.1]{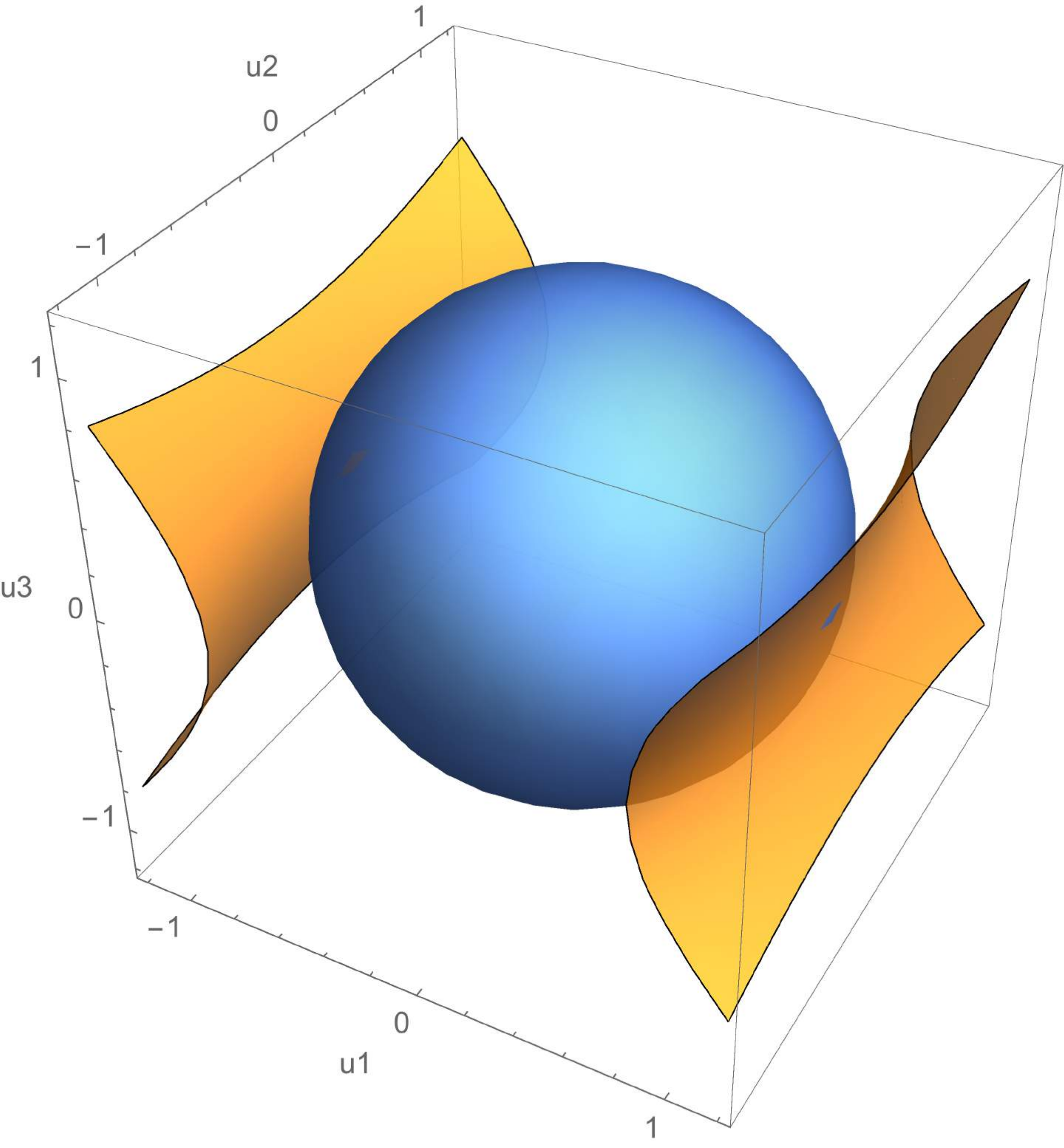}  }
\end{center}
\caption{Intersections of the Casimir function (\ref{Con2E}) and the unitary sphere when $e_0=e_i$, $i=1,2,3$.}
\label{intersec}
\end{figure}

For $b \neq 0$ the Casimir $H$ does not coincide anymore with $h$ and its geometries are discussed next.


\subsubsection{Asymmetric cases: $D_2 \subset SO(3)$}

As discussed in \cite{Patera}, for the asymmetric case, the Dihedral $D_2$ group is a subgroup of  $SO(3)$. 
We use the notation $D_2 \subset SO(3)$ to emphasize this property. In order to determine the geometries of 
the Casimir function $H$ it is important to know the relation between the energy parameter $e_0$ and the inertia 
parameter $e_2$. We have as in section \ref{DimEuler}, two different cases.

\begin{itemize}
\item Case  $e_2 < e_0 < e_1$
\end{itemize}

In this first case, the dimensionless energy parameter $e_0$ lies between the values of $e_1$ and $e_2$. To 
obtain the different geometries of the $H$ surface, we allow the free parameter $b$ to take every value in the real 
line. This gives origin to 9 different generic cases. For instance, if $-\infty < b<-ae_1$, the three coefficients $ae_i+b$ 
are negative (see table \ref{table1Comb}) as well as the quantity $ae_0+b$. The resulting surface is an ellipsoid 
whose larger axis is in the direction $u_1$, whereas the shorter axis is in direction $u_3$. In the limit situation 
$b \rightarrow - \infty$ the $H$ surface degenerates into an $S^2$ sphere. When the parameter takes the exact value 
$b=-ae_1$ the $H$ surface becomes an elliptic cylinder with symmetry axis in the direction $u_1$. 
If now the values of the parameter are in the interval $-ae_1 < b < -ae_0$, the $H$ surface becomes an elliptic 
hyperboloid of one sheet around axis $u_1$. The different surfaces $H$ obtained for the different values of the 
parameter $b$ are summarized in table \ref{table1Comb}. 

\begin{table}[htb]  
\begin{tabular}{|c | c | c | c | c |} 
\hline
$ae_1+b$ & $ae_0+b$ & $ae_2+b$ & $ae_3+b$ & $H$ surface \\
\hline 
\hline
 $< 0$ & $<0$ & $<0$ & $<0$ &  elliptic ellipsoid with larger axis in $u_1$ and shorter axis in $u_3$\\ 
 $=0$ &  $< 0$ &  $< 0$ & $<0$ &  elliptic cylinder around $u_1$\\
 $>0$ & $<0$ &  $<0$ &  $<0$ &  elliptic hyperboloid of one sheet around $u_1$\\
 $>0$ & $=0$ &  $<0$ &  $<0$ &   elliptic cone around $u_1$ \\
 $>0$ & $>0$ &  $<0$ &  $<0$ &elliptic hyperboloid of two sheets around $u_1$ \\ 
 $>0$ & $>0$ &  $=0$ &  $<0$ &  hyperbolic cylinder around $u_2$\\
 $>0$ & $>0$ &  $> 0$ &  $<0$ &  elliptic hyperboloid of one sheet around $u_3$\\
 $>0$ & $>0$ &  $> 0$ &  $=0$ &  elliptic cylinder around $u_3$ \\
 $>0$ & $>0$ &  $> 0$ &  $>0$ &  elliptic ellipsoid with larger axis in $u_3$ and shorter axis in $u_1$\\
\hline
\end{tabular}
\caption{Classification of the nine different geometries for the Casimir function $H$ (\ref{DefH1}), in the 
asymmetric cases satisfying $e_2 < e_0 < e_1 $. For each of these cases the Lie algebra of the Hamiltonian vector 
fields associated to the coordinates $u_i$'s is of Bianchi IX type.}\label{table1Comb}
\end{table}

It is important to stress that independently of the geometry of $H$ the important fact is that the intersection of the 
surfaces $H$ and $N$ is the same. In other words, the dynamics of the asymmetric rigid body, which takes place 
on the intersection of the two Casimir surfaces according to the Euler equations (\ref{EulerHN}), is invariant 
under a gauge transformation of the form (\ref{GaugeIX}), and the geometrical meaning of the transformation 
consists in changing the geometry of the Casimir $H$.

\begin{itemize}
\item Case  $e_3 < e_0 < e_2 $
\end{itemize}
This case is very similar to the previous one, but this time, because $e_0$ belongs to a different interval, the 
intersections of the Casimir functions $H$ and $N$  occur in different regions. In the previous case the intersections 
occur in the upper region respect to the separatrix,  and are given by elliptical curves on the sphere around the axis 
$u_3$, whereas in this case the intersections occur below the separatrix and are given by elliptical curves on the 
sphere around the axis $u_1$. In table \ref{table2Comb} we summarize the geometries for the different possible 
values of the parameter $b$.

\begin{table}[htb]  
\begin{tabular}{|c | c | c | c | c |} 
\hline
$ae_1+b$ & $ae_2+b$ & $ae_0+b$ & $ae_3+b$ & $H$ Surface \\
\hline 
\hline
 $ < 0$ & $<0$ & $<0$ & $<0$ & elliptic ellipsoid with larger axis in $u_1$ and shorter axis in $u_3$ \\ 
 $=0$ &  $< 0$ &  $< 0$ & $<0$ &  elliptic cylinder around $u_1$ \\
 $>0$ & $<0$ &  $<0$ &  $<0$ &  elliptic hyperboloid of one sheet around $u_1$ \\
 $>0$ & $=0$ &  $<0$ &  $<0$ &  hyperbolic cylinder around $u_2$ \\
 $>0$ & $>0$ &  $<0$ &  $<0$ & elliptic hyperboloid of two sheets around $u_3$\\ 
 $>0$ & $>0$ &  $=0$ &  $<0$ &  elliptic cone around $u_3$ \\
 $>0$ & $>0$ &  $> 0$ &  $<0$ &  elliptic hyperboloid of one sheet  around $u_3$\\
 $>0$ & $>0$ &  $> 0$ &  $=0$ &   elliptic cylinder around $u_3$\\
 $>0$ & $>0$ &  $> 0$ &  $>0$ &   elliptic ellipsoid with larger axis in $u_3$ and shorter axis in $u_1$ \\
\hline
\end{tabular}
\caption{Classification of the nine different geometries for the Casimir function $H$ (\ref{DefH1}), in the 
asymmetric cases satisfying $e_3 < e_0 < e_2$. For each of these cases the Lie algebra of the Hamiltonian vector 
fields associated to the coordinates $u_i$'s is of Bianchi IX type.}\label{table2Comb}
\end{table}
Finally notice that in transformations (\ref{GaugeIX}) we have considered that $a>0$. If instead we consider group 
elements such that $a<0$, the gauge transformation changes the order of the inertia parameters from 
$e_3<e_2<e_1$ to $e_1<e_2<e_3$ and therefore to an interchange in the solutions of the modulus of the elliptic 
functions $k_1^2 \leftrightarrow k_2^2$ (see table \ref{orderings}).


\subsubsection{Symmetric cases: $SO(2) \subset SO(3)$}

As limiting cases of the Bianchi IX asymmetric rigid bodies we obtain either the oblate (prolate) Bianchi IX symmetric 
rigid bodies. For these cases $SO(2)$ is a subgroup of the $SO(3)$ Lie algebra (\ref{SO3})  and using the notation 
of \cite{Patera} we will denote both cases as $SO(2) \subset SO(3)$.
As discussed in section \ref{Gauge}, the symmetric cases are obtained by setting $e_1=e_2$ (oblate) and 
$e_2=e_3$ (prolate). 
Taking the oblate limit in table \ref{table2Comb} we obtain seven different geometries for the surface $H$ which are 
summarized in the following table.

\begin{table}[htb]  
\begin{tabular}{|c | c | c | c |} 
\hline
$ae_1+b$  & $ae_0+b$ & $ae_3+b$ & $H$ Surface \\
\hline 
\hline
 $ < 0$ & $<0$ & $<0$ &  oblate ellipsoid  with larger axis in $u_1$\\ 
 $=0$ &  $< 0$ & $<0$ &  parallel $u_1 u_2$ planes \\
 $>0$ & $<0$ &  $<0$ &  circular hyperboloid of two sheets around $u_3$ \\
 $>0$ & $=0$ &  $<0$ &  two sheet circular cone around $u_3$\\
 $>0$ & $>0$ &  $<0$ & circular hyperboloid of one sheet around $u_3$\\ 
 $>0$ &  $> 0$ &  $=0$ &   circular cylinder around $u_3$\\
 $>0$ & $>0$ &  $> 0$ &  oblate ellipsoid with larger axis in $u_3$\\
\hline
\end{tabular}
\caption{Classification of the seven different geometries for the Casimir function $H$ (\ref{DefH1}), in the 
oblate symmetric case. For each of these cases the Lie algebra of the Hamiltonian vector 
fields associated to the coordinates $u_i$'s is of Bianchi IX type.}\label{table2Symmetric}
\end{table}

In a similar way, the prolate symmetric cases are obtained as limiting cases of the classification in table 
\ref{table1Comb}.  There are again seven different geometries and they differ respect to the ones in the oblate case 
in the orientation of the axial axis of symmetry. Because to obtain them is straightforward we do not go into details.


\subsection{ Cases: $c\neq 0$ and $d = 0$}\label{Cased0}

A generic $SL(2,\mathbb{R})$ group element of this case has the form
\begin{equation}\label{gauge2}
g= \left( 
\begin{array}{cc}
a & b \\
-1/b & 0
\end{array}
\right) \hspace{0.3cm} \mbox{with}  \hspace{0.3cm}  b  \in \mathbb{R} \setminus  \{0\}. 
\end{equation} 
The case $a=0$ only interchanges the role of the Casimir functions $h$ and $l$ an therefore is of Bianchi type IX.
Here without loosing generality we assume that $a>0$. In this case the Lie algebra (\ref{GeneralLie}) reduces to 
the form 
\begin{equation}\label{bianchi8}
[X_{u_1},X_{u_2}]=-\frac{1}{b}e_3 X_{u_3}, \hspace{0.5cm} 
[X_{u_2},X_{u_3}]=-\frac{1}{b}e_1 X_{u_1}, \hspace{0.5cm}
[X_{u_3},X_{u_1}]=-\frac{1}{b}e_2 X_{u_2}.
\end{equation}
The Casimir function $H$ has the same form as (\ref{DefH1}) and therefore its geometries are given according to 
tables \ref{table1Comb} and \ref{table2Comb}, whereas the Casimir function $N$ in this case is proportional to $h$: 
$N=-h/b$, and therefore its geometries are classified according to tables \ref{table1} and \ref{table2}. As we discuss  
next the group elements (\ref{gauge2}) act as gauge transformations in two inequivalent 
Lie algebras. The first one corresponds to the Bianchi VIII or $SO(2,1)$ Lie algebra whereas the second one to 
the Bianchi VI$_0$ or $ISO(1,1)$ algebra. The difference of these two algebras is due to the value of the $e_2$ 
inertia parameter. If $e_2 \neq 0$ we are in the Bianchi VIII case and if $e_2=0$ we have the Bianchi VI$_0$ case. 
We stress that the generators in both algebras describe in general an asymmetric body. Strictly speaking the 
system does not corresponds to a rigid body anymore because the Casimir function $N$  is not longer $S^2$. 
Sometimes in the literature these systems are called generalized rigid bodies \cite{Iwai2010501}.  Finally we 
comment that as in the previous subsection, taking $a<0$ in (\ref{gauge2}) produces a different order in the inertia 
parameters.  

\subsubsection{Bianchi VIII cases}

The Lie algebra (\ref{bianchi8}) is in the Bianchi VIII class if $e_2 \neq 0$ as we can easily see. First remember 
that $e_1>0$, $e_3 =- |e_3| <0$ and rewrite $e_2 \equiv \epsilon |e_2|$ where $\epsilon \equiv$ sign$(e_2)$ since 
$e_2$ can be either positive or negative. If we redefine the generators (\ref{BasicGen}) in the way
\begin{equation}\label{DefXVIII}
X_{u_1} \equiv -\frac{1}{b} \sqrt{|e_2||e_3|} Y_{u_1}, \hspace{0.5cm} 
X_{u_2} \equiv -\frac{1}{b} \sqrt{e_1 |e_3|} Y_{u_2}, \hspace{0.5cm} 
X_{u_3} \equiv -\frac{1}{b} \sqrt{e_1 |e_2|} Y_{u_3},
\end{equation}
we obtain the $SO(2,1)$ Lie algebra
\begin{equation}
[Y_{u_1},Y_{u_2}]= -Y_{u_3}, \hspace{0.5cm} 
[Y_{u_2},Y_{u_3}]= + Y_{u_1}, \hspace{0.5cm}
[Y_{u_3},Y_{u_1}]= \epsilon Y_{u_2}.
\end{equation}
The symmetric limits can be taken directly in expressions (\ref{DefXVIII}) by setting $e_1=e_2$ (oblate) and 
$e_2=e_3$ (prolate).

\subsubsection{Bianchi VI$_0$ case}

The second inequivalent Lie algebra emerges when $e_2=0 \Rightarrow e_1=-e_3$. In this case the geometry 
of the Casimir function $N$ corresponds to a parabolic cylinder, according to tables \ref{table1} and \ref{table2}. 
Redefining the generators (\ref{BasicGen}) as $X_{u_i} \equiv -\sqrt{e_1} \, Y_{u_i}/b$, for ${i=1,3}$ and 
$X_{u_2} \equiv - e_1 \, Y_{u_2}/b$,  the Lie algebra is of  Bianchi VI$_0$ type
\begin{equation}
[Y_{u_1},Y_{u_2}]= -Y_{u_3}, \hspace{0.5cm} 
[Y_{u_2},Y_{u_3}]=+Y_{u_1}, \hspace{0.5cm}
[Y_{u_3},Y_{u_1}]=0.
\end{equation}
Notice in this case the expression of the vector fields (\ref{BasicGen}) reduce to
\begin{equation}
X_{u_1}= - u_3 \partial_2, \hspace{0.5cm}
X_{u_2}= u_1 \partial_3+ u_3 \partial_1, \hspace{0.5cm}
X_{u_3}=  - u_1 \partial_2,
\end{equation}
where the value of the modulus in the Jacobi elliptic functions in the partials (\ref{Killings}) are $k_1^2=k_2^2=1/2$.
There is not symmetric limit for this case.


\subsection{Case $c \neq 0$, $d \neq 0$}

In this case both Casimir functions are given by linear combinations of $h$ and $l$. $H$ is given by expression 
(\ref{DefH1}) whereas $N$ is given by a similar expression
\begin{equation}
N=\frac{1}{2} \left[ (ce_1 + d ) u_1^2 + (ce_2+d) u_2^2 + (ce_3 +d) u_3^2 \right]  - \frac{1}{2}(ce_0+d) =0.
\end{equation}
For these gauge transformations the Lie algebra (\ref{GeneralLie}) can be rewritten as
\begin{equation}\label{LastCases}
[Y_{u_1},Y_{u_2}]= \epsilon_3 Y_{u_3}, \hspace{0.5cm} 
[Y_{u_2},Y_{u_3}]= \epsilon_1 Y_{u_1}, \hspace{0.5cm}
[Y_{u_3},Y_{u_1}]= \epsilon_2 Y_{u_2}.
\end{equation}
It turns out that the general asymmetric case contains seven cases which belong to four different type A algebras in 
the Bianchi classification, namely Bianchi IX, VIII, VII$_0$ and VI$_0$. Without loosing generality we assume that 
$c>0$. In table \ref{Bianchicd} we summarize all the different cases, which we analyze below.

\begin{table}[htb]  
\begin{tabular}{| c | c | c | c | c | c | c | c |} 
\hline
Situation & $ce_1+d$ & $ce_2+d$ & $ce_3+d$ & $\epsilon_1$ & $\epsilon_2$ & $\epsilon_3$ & Bianchi type \\
\hline 
\hline
1 & $< 0$& $<0$ & $<0$ & $-1$ &  $-1$ & $-1$ & IX  \\ 
2 & $=0$ & $< 0$& $< 0$ & $0$ & $-1$ & $-1$ & VII$_0$ \\
3 & $>0$ & $<0$ & $<0$ &  $+1$ & $-1$ & $-1$ & VIII \\
4 & $>0$ & $=0$ & $<0$ &  $+1$ &  $0$ & $-1$ & VI$_0$ \\
5 & $>0$ & $>0$ & $<0$ &  $+1$ & $+1$ & $-1$ & VIII \\ 
6 & $>0$ & $>0$ & $=0$ &  $+1$ & $+1$ & $0$ & VII$_0$  \\
7 & $>0$ & $>0$ & $> 0$ & $+1$ & $+1$ & $+1$ & IX\\
\hline
\end{tabular}
\caption{Inequivalent Lie algebras for $SL(2\mathbb{R})$ gauge transformations with $c\neq 0$ and $d \neq 0$. 
All the algebras are of type A in the Bianchi classification, and describe an asymmetric rigid body.}\label{Bianchicd}
\end{table}

\subsubsection{Asymmetric Bianchi IX and VIII cases}

Proceeding in a similar way to the analysis performed in section \ref{Cased0}, we write the three coefficients 
$ce_i+d \equiv \epsilon_i |ce_i+d| \neq 0$, where $\epsilon_i=$sign$(ce_i+d)$. As a 
consequence we define the generators $Y_i$'s trough the relations
\begin{equation}\label{Bianchi9}
X_{u_1} \equiv  \sqrt{|c e_3+d||c e_2+d|} Y_{u_1}, \hspace{0.5cm} 
X_{u_2} \equiv  \sqrt{|c e_3+d||c e_1+d|} Y_{u_2}, \hspace{0.5cm} 
X_{u_3} \equiv  \sqrt{|c e_2+d||c e_1+d|} Y_{u_3},
\end{equation}
which drive directly to equation (\ref{LastCases}). Situations 1, 3, 5 and 7 in table \ref{Bianchicd} are of this form. 

\subsubsection{Asymmetric Bianchi VII$_0$ and VI$_0$ cases}

\begin{itemize}
\item Situation 2: $d=-c e_1$, (Bianchi type VII$_0$ or $ISO(2)$).
\end{itemize}
According to table \ref{Bianchicd}, this case is obtained when the coefficients of the $SL(2,\mathbb{R})$ 
transformation satisfy the relation $d=-c e_1$. In this case the Lie algebra (\ref{LastCases}) is of Bianchi type VII$_0$ 
or $ISO(2)$. Without lossing generality we can assume that $c>0$ and the suitable redefinition of generators is 
given by
\begin{equation}
X_{u_1} \equiv  c \sqrt{ (e_1-e_3)(e_1-e_2)} Y_{u_1}, \hspace{0.5cm} 
X_{u_2} \equiv c  \alpha Y_{u_2}, \hspace{0.5cm} 
X_{u_3} \equiv  c \alpha \sqrt{ \frac{e_1-e_2}{e_1-e_3}} Y_{u_3}.
\end{equation}
Notice that additionally to the constant $c$, the generators $X_{u_2}$ and $X_{u_3}$ can be multiplied by an 
arbitrary constant $\alpha$.

\begin{itemize}
\item Situation 4: $d=-c e_2 \neq 0$  (Bianchi VI$_0$ or $ISO(1,1)$).
\end{itemize}
In this case the suitable redefinition of generators is given by
\begin{equation}
X_{u_1} \equiv  c \alpha \sqrt{ \frac{e_2-e_3}{e_1-e_2}} Y_1, \hspace{0.5cm} 
X_{u_2} \equiv c \sqrt{(e_1-e_2)(e_2-e_3)}  Y_2, \hspace{0.5cm} 
X_{u_3} \equiv  c \alpha  Y_3.
\end{equation}

\begin{itemize}
\item Situation 6: $d=-c e_3$ (Bianchi VII$_0$ or $ISO(2)$).
\end{itemize}
In this case the edefinition of generators is given as
\begin{equation}\label{Bianchi7}
X_{u_1} \equiv  c \alpha    Y_1, \hspace{0.5cm} 
X_{u_2} \equiv c \alpha  \sqrt{\frac{e_1-e_3}{e_2-e_3}} Y_2, \hspace{0.5cm} 
X_{u_3} \equiv  c \sqrt{(e_1-e_3)(e_2-e_3)}  Y_3.
\end{equation}


\subsubsection{Oblate symmetric limits ($e_1=e_2$)}

Regarding the oblate symmetric limits ($e_1=e_2$), we can see from table \ref{Bianchicd} that they can be naturally  
obtained in the situations 1, 5, 6, 7, because in this case the signs of both $ce_1+d$ and $ce_2+d$ are the same.  
The cases 2, 3 and 4 all degenerate into one single case which is characterized by 
having $ce_1+d=ce_2+d=0$ and whose algebra is of Bianchi II type and commonly known as the Heinseberg  
algebra Heis$_3$ \cite{Gilmore:2008zz}
\begin{table}[htb]  
\begin{tabular}{| c | c | c | c | c | c | c |} 
\hline
Situation &  $ce_1+d$ &  $ce_3+d$  & Bianchi type & $X_{u_1}$ & $X_{u_2}$ & $X_{u_3}$ \\
\hline 
\hline
1 & $<0$ & $<0$ & IX  &  $ \sqrt{|c e_3+d||c e_1+d|} Y_{u_1} $ &  
$ \sqrt{|c e_3+d||c e_1+d|} Y_{u_2}$ &  $ |c e_1+d|  Y_{u_3}$ \\ 
8 &$ =0$ & $< 0$ & II &$ Y_{u_1} $ &  $ Y_{u_2}$ & $3|d|  Y_{u_3}/2$ \\
5 & $>0$ & $<0$ &  VIII   &   $ \sqrt{|c e_3+d||c e_1+d|} Y_{u_1} $ &  
$ \sqrt{|c e_3+d||c e_1+d|} Y_{u_2}$ &  $ |c e_1+d|  Y_{u_3}$  \\ 
6 & $>0$ & $=0$ &  VII$_0$  &  $ Y_{u_1} $ &  
$ Y_{u_2}$ &  $ |c e_1+d| Y_{u_3}$ \\
7 &$>0$ & $> 0$ & IX  &   $ \sqrt{|c e_3+d||c e_1+d|} Y_{u_1} $ &  
$ \sqrt{|c e_3+d||c e_1+d|} Y_{u_2}$ &  $ |c e_1+d|  Y_{u_3}$\\
\hline
\end{tabular}
\caption{Oblate symmetric cases ($e_1=e_2$) and the classification of the Lie-Poisson algebra of the Hamiltonian 
vector fields associated with the coordinates $u_i$.} \label{BianchiSymm}
\end{table}

\subsubsection{Prolate symmetric limits ($e_2=e_3$)}

A similar result is obtained for the prolate limit ($e_2=e_3$). For completeness we include the possibilities in the 
following table, but we do not going into details as the algebras resulting from taking this limit are the same as 
in the $e_1 = e_2$ case.
\begin{table}[h!]  
\begin{tabular}{| c | c | c | c | } 
\hline
Situation & $ce_1+d$ & $ce_3+d$ & Bianchi type \\
\hline 
\hline
1 & $< 0$& $<0$ & IX  \\ 
2 & $= 0$& $<0$ & VII$_0$  \\ 
3 & $>0$ & $<0$ &  VIII \\ 
8 & $>0$ & $=0$ &  II  \\
7 & $>0$ & $>0$ & IX\\
\hline
\end{tabular}
\caption{Prolate symmetric cases ($e_2=e_3$) and the classification of the Lie-Poisson algebra of the Hamiltonian 
vector fields associated with the coordinates $u_i$.}\label{BianchiSymm}
\end{table}


\section{Conclusions}\label{conclusions}

In this contribution we have fully classified the $SL(2,\mathbb{R})$ gauge transformations in the rigid body, using 
the generalized dynamics of Nambu. These gauge transformations are geometrical transformations of the Casimir 
functions that keep invariant its geometrical intersections once the Hamiltonian vector fields associated to the 
coordinates (the components of angular momentum in this case) satisfy a specific three dimensional Lie algebra. 
Some general results were previously reported in the literature \cite{Marsden1,Iwai2010501}, and here 
we have constructed explicitly all the inequivalent Bianchi type A Lie algebras that are possible for this system. 
In particular we have worked out and classified the Hamiltonian vector fields for all the Lie algebras, something that 
has not been fully addressed in the literature.  Our analysis is classical  and  can be considered as a 
first step in the direction to generalize the quantum results obtained for the symmetrical rigid bodies 
\cite{,Iwai2010501}, to the asymmetrical ones.  In fact there are some recent results in the study of the quantum 
spectrum of the rotational motion of asymmetric molecules 
\cite{Pina1999159,Pina,LeyKoo,MendezFragoso2015115,ley2015rotations}, that could be generalized 
to the extended rigid body problem, and could bring some light in the quantization of the system using the Nambu 
brackets for the asymmetric cases.

Although the rigid body problem has been extensively studied over the years,  from the classical mechanics point of 
view there are still some interesting issues that could be addressed. 
For mentioning some of them, it would be very interesting to stablish the nature of the generating function of the 
canonical transformations from the coordinates (\ref{spheroconal}) used here and the angle-action coordinates for 
the extended rigid bodies. A similar study was recently performed for the simple pendulum 
\cite{Brizard2013511}, and because the dynamics of the rigid body reduces to the one of the simple 
pendulum when the Lie algebra is of Bianchi type VII$_0$ or $ISO(2)$, it must be possible to perform such an  
analysis.  Another possibility is to study the $SL(3,\mathbb{R})$ canonical transformations of the rigid 
body in the context of the generalized bi-Hamiltonian dynamics and determine which canonical transformations 
generate the different ordering in the inertia parameters, because as we have discussed in this contribution, the  
$SL(2,\mathbb{R})$ gauge transformations give account only of the ones that interchange the parameters 
$e_1 \leftrightarrow e_3$. This analysis should be possible because the solutions for the asymmetric 
cases are given in terms of Jacobi elliptical functions and they incorporate in a natural way the transformations 
to generate the six different case of table \ref{orderings}. For a recent review in the case of the simple pendulum 
see \cite{Linares}. A further very interesting analysis is find the relation between the different gauge 
transformations studied in this paper with the role these same transformations play in the context of the generalized 
Hamilton-Jacobi theory of Nambu Mechanics recently studied in \cite{Yoneya:2016abn}.


\acknowledgments 
The work of M. de la C. and N. G. is supported by the Ph.D. scholarship program of the Universidad 
Aut\'{o}noma Metropolitana. The work of R. L. is  partially supported from CONACyT Grant No. 237351 
``Implicaciones f\'{i}sicas de la estructura del espacio-tiempo".

\bibliography{notes}

\begin{thebibliography}{34}
\expandafter\ifx\csname natexlab\endcsname\relax\def\natexlab#1{#1}\fi
\expandafter\ifx\csname bibnamefont\endcsname\relax
  \def\bibnamefont#1{#1}\fi
\expandafter\ifx\csname bibfnamefont\endcsname\relax
  \def\bibfnamefont#1{#1}\fi
\expandafter\ifx\csname citenamefont\endcsname\relax
  \def\citenamefont#1{#1}\fi
\expandafter\ifx\csname url\endcsname\relax
  \def\url#1{\texttt{#1}}\fi
\expandafter\ifx\csname urlprefix\endcsname\relax\def\urlprefix{URL }\fi
\providecommand{\bibinfo}[2]{#2}
\providecommand{\eprint}[2][]{\url{#2}}

\bibitem[{\citenamefont{Nambu}(1973)}]{Nambu:1973qe}
\bibinfo{author}{\bibfnamefont{Y.}~\bibnamefont{Nambu}},
  \bibinfo{journal}{Phys. Rev.} \textbf{\bibinfo{volume}{D7}},
  \bibinfo{pages}{2405} (\bibinfo{year}{1973}).

\bibitem[{\citenamefont{Curtright and
  Zachos}(2003{\natexlab{a}})}]{Curtright:2002fd}
\bibinfo{author}{\bibfnamefont{T.}~\bibnamefont{Curtright}} \bibnamefont{and}
  \bibinfo{author}{\bibfnamefont{C.~K.} \bibnamefont{Zachos}},
  \bibinfo{journal}{Phys. Rev.} \textbf{\bibinfo{volume}{D68}},
  \bibinfo{pages}{085001} (\bibinfo{year}{2003}{\natexlab{a}}),
  \eprint{hep-th/0212267}.

\bibitem[{\citenamefont{Curtright and
  Zachos}(2003{\natexlab{b}})}]{Curtright:2003fn}
\bibinfo{author}{\bibfnamefont{T.}~\bibnamefont{Curtright}} \bibnamefont{and}
  \bibinfo{author}{\bibfnamefont{C.~K.} \bibnamefont{Zachos}},
  \bibinfo{journal}{AIP Conf. Proc.} \textbf{\bibinfo{volume}{672}},
  \bibinfo{pages}{165} (\bibinfo{year}{2003}{\natexlab{b}}),
  \eprint{hep-th/0303088}.

\bibitem[{\citenamefont{Minic}(1999)}]{Minic:1999js}
\bibinfo{author}{\bibfnamefont{D.}~\bibnamefont{Minic}} (\bibinfo{year}{1999}),
  \eprint{hep-th/9909022}.

\bibitem[{\citenamefont{Ho and Matsuo}(2016)}]{Ho:2016hob}
\bibinfo{author}{\bibfnamefont{P.-M.} \bibnamefont{Ho}} \bibnamefont{and}
  \bibinfo{author}{\bibfnamefont{Y.}~\bibnamefont{Matsuo}}, in
  \emph{\bibinfo{booktitle}{{Nambu Memorial Symposium Osaka, Japan, September
  29, 2015}}} (\bibinfo{year}{2016}), vol. \bibinfo{volume}{2016}, p.
  \bibinfo{pages}{06A104}, \eprint{1603.09534}.

\bibitem[{\citenamefont{Yoneya}(2016{\natexlab{a}})}]{Yoneya:2016aja}
\bibinfo{author}{\bibfnamefont{T.}~\bibnamefont{Yoneya}}
  (\bibinfo{year}{2016}{\natexlab{a}}), \eprint{1612.08513}.

\bibitem[{\citenamefont{Naruki and Tarama}(2011)}]{Naruki2011S170}
\bibinfo{author}{\bibfnamefont{I.}~\bibnamefont{Naruki}} \bibnamefont{and}
  \bibinfo{author}{\bibfnamefont{D.}~\bibnamefont{Tarama}},
  \bibinfo{journal}{Differential Geometry and its Applications}
  \textbf{\bibinfo{volume}{29, Supplement 1}}, \bibinfo{pages}{S170 }
  (\bibinfo{year}{2011}).

\bibitem[{\citenamefont{Holm and Marsden}(1991)}]{Marsden1}
\bibinfo{author}{\bibfnamefont{D.~D.} \bibnamefont{Holm}} \bibnamefont{and}
  \bibinfo{author}{\bibfnamefont{J.~E.} \bibnamefont{Marsden}},
  \bibinfo{journal}{In Honor of J.-M. Souriau} pp. \bibinfo{pages}{1--15}
  (\bibinfo{year}{1991}).

\bibitem[{\citenamefont{Minic and Tze}(2002)}]{Minic:2002pd}
\bibinfo{author}{\bibfnamefont{D.}~\bibnamefont{Minic}} \bibnamefont{and}
  \bibinfo{author}{\bibfnamefont{H.~C.} \bibnamefont{Tze}},
  \bibinfo{journal}{Phys. Lett.} \textbf{\bibinfo{volume}{B536}},
  \bibinfo{pages}{305} (\bibinfo{year}{2002}), \eprint{hep-th/0202173}.

\bibitem[{\citenamefont{Iwai and Tarama}(2010)}]{Iwai2010501}
\bibinfo{author}{\bibfnamefont{T.}~\bibnamefont{Iwai}} \bibnamefont{and}
  \bibinfo{author}{\bibfnamefont{D.}~\bibnamefont{Tarama}},
  \bibinfo{journal}{Differential Geometry and its Applications}
  \textbf{\bibinfo{volume}{28}}, \bibinfo{pages}{501 } (\bibinfo{year}{2010}).

\bibitem[{\citenamefont{Kramers and Ittmann}(1929)}]{Ittmann}
\bibinfo{author}{\bibfnamefont{H.~A.} \bibnamefont{Kramers}} \bibnamefont{and}
  \bibinfo{author}{\bibfnamefont{G.~P.} \bibnamefont{Ittmann}},
  \bibinfo{journal}{Zs. f. Phys.} \textbf{\bibinfo{volume}{53}},
  \bibinfo{pages}{553} (\bibinfo{year}{1929}).

\bibitem[{\citenamefont{Lukac and Smorodinskii}(1970)}]{LukacSmoro}
\bibinfo{author}{\bibfnamefont{I.}~\bibnamefont{Lukac}} \bibnamefont{and}
  \bibinfo{author}{\bibfnamefont{A.}~\bibnamefont{Smorodinskii}},
  \bibinfo{journal}{Soviet Phys. JETP} \textbf{\bibinfo{volume}{30}},
  \bibinfo{pages}{728} (\bibinfo{year}{1970}).

\bibitem[{\citenamefont{Pi\~{n}a}(1996)}]{Pinna}
\bibinfo{author}{\bibfnamefont{E.}~\bibnamefont{Pi\~{n}a}},
  \emph{\bibinfo{title}{{Din\'amica de Rotaciones}}}
  (\bibinfo{publisher}{Colecci\'on CBI, Universidad Aut\'onoma Metropolitana},
  \bibinfo{year}{1996}).

\bibitem[{\citenamefont{Pi\~{n}a}(1999)}]{Pina1999159}
\bibinfo{author}{\bibfnamefont{E.}~\bibnamefont{Pi\~{n}a}},
  \bibinfo{journal}{Journal of Molecular Structure: \{THEOCHEM\}}
  \textbf{\bibinfo{volume}{493}}, \bibinfo{pages}{159 } (\bibinfo{year}{1999}).

\bibitem[{\citenamefont{Bianchi}(2001)}]{Bianchi2001}
\bibinfo{author}{\bibfnamefont{L.}~\bibnamefont{Bianchi}},
  \bibinfo{journal}{General Relativity and Gravitation}
  \textbf{\bibinfo{volume}{33}}, \bibinfo{pages}{2171} (\bibinfo{year}{2001}).

\bibitem[{\citenamefont{Vald\'{e}s and Pi\~{n}a}(2006)}]{Pina}
\bibinfo{author}{\bibfnamefont{M.~T.} \bibnamefont{Vald\'{e}s}}
  \bibnamefont{and} \bibinfo{author}{\bibfnamefont{E.}~\bibnamefont{Pi\~{n}a}},
  \bibinfo{journal}{Rev. Mex. Fis.} \textbf{\bibinfo{volume}{52}},
  \bibinfo{pages}{220} (\bibinfo{year}{2006}).

\bibitem[{\citenamefont{M\'{e}ndez-Fragoso and Ley-Koo}(2011)}]{LeyKoo}
\bibinfo{author}{\bibfnamefont{R.}~\bibnamefont{M\'{e}ndez-Fragoso}}
  \bibnamefont{and} \bibinfo{author}{\bibfnamefont{E.}~\bibnamefont{Ley-Koo}},
  \bibinfo{journal}{Advances in Quantum Chemistry}
  \textbf{\bibinfo{volume}{62}}, \bibinfo{pages}{137} (\bibinfo{year}{2011}).

\bibitem[{\citenamefont{M\'{e}ndez-Fragoso and
  Ley-Koo}(2015)}]{MendezFragoso2015115}
\bibinfo{author}{\bibfnamefont{R.}~\bibnamefont{M\'{e}ndez-Fragoso}}
  \bibnamefont{and} \bibinfo{author}{\bibfnamefont{E.}~\bibnamefont{Ley-Koo}},
  in \emph{\bibinfo{booktitle}{Concepts of Mathematical Physics in Chemistry: A
  Tribute to Frank E. Harris - Part A}}, edited by
  \bibinfo{editor}{\bibfnamefont{J.~R.} \bibnamefont{Sabin}} \bibnamefont{and}
  \bibinfo{editor}{\bibfnamefont{R.}~\bibnamefont{Cabrera-Trujillo}}
  (\bibinfo{publisher}{Academic Press}, \bibinfo{year}{2015}),
  vol.~\bibinfo{volume}{71} of \emph{\bibinfo{series}{Advances in Quantum
  Chemistry}}, pp. \bibinfo{pages}{115 -- 152}.

\bibitem[{\citenamefont{Ley-Koo}(2015)}]{ley2015rotations}
\bibinfo{author}{\bibfnamefont{E.}~\bibnamefont{Ley-Koo}}, in
  \emph{\bibinfo{booktitle}{Journal of Physics: Conference Series}}
  (\bibinfo{organization}{IOP Publishing}, \bibinfo{year}{2015}), vol.
  \bibinfo{volume}{597}, p. \bibinfo{pages}{012055}.

\bibitem[{\citenamefont{Reiche}(1918)}]{Reiche}
\bibinfo{author}{\bibfnamefont{F.}~\bibnamefont{Reiche}},
  \bibinfo{journal}{Physik. Z.} \textbf{\bibinfo{volume}{19}},
  \bibinfo{pages}{394} (\bibinfo{year}{1918}).

\bibitem[{\citenamefont{King}(1947)}]{King}
\bibinfo{author}{\bibfnamefont{G.~W.} \bibnamefont{King}},
  \bibinfo{journal}{The Journal of Chemical Physics}
  \textbf{\bibinfo{volume}{15}}, \bibinfo{pages}{820} (\bibinfo{year}{1947}).

\bibitem[{\citenamefont{Spence}(1959)}]{spence}
\bibinfo{author}{\bibfnamefont{R.~D.} \bibnamefont{Spence}},
  \bibinfo{journal}{Am. J. Phys.} \textbf{\bibinfo{volume}{27}},
  \bibinfo{pages}{329} (\bibinfo{year}{1959}).

\bibitem[{\citenamefont{David and Holm}(1992)}]{David1992}
\bibinfo{author}{\bibfnamefont{D.}~\bibnamefont{David}} \bibnamefont{and}
  \bibinfo{author}{\bibfnamefont{D.~D.} \bibnamefont{Holm}},
  \bibinfo{journal}{Journal of Nonlinear Science} \textbf{\bibinfo{volume}{2}},
  \bibinfo{pages}{241} (\bibinfo{year}{1992}).

\bibitem[{\citenamefont{Patera and Winternitz}(1973)}]{Patera}
\bibinfo{author}{\bibfnamefont{J.}~\bibnamefont{Patera}} \bibnamefont{and}
  \bibinfo{author}{\bibfnamefont{P.}~\bibnamefont{Winternitz}},
  \bibinfo{journal}{Journal of Mathematical Physics}
  \textbf{\bibinfo{volume}{14}}, \bibinfo{pages}{1130} (\bibinfo{year}{1973}).

\bibitem[{\citenamefont{Gilmore}(2008)}]{Gilmore:2008zz}
\bibinfo{author}{\bibfnamefont{R.}~\bibnamefont{Gilmore}},
  \emph{\bibinfo{title}{{Lie groups, physics, and geometry: An introduction for
  physicists, engineers and chemists}}} (\bibinfo{year}{2008}).

\bibitem[{\citenamefont{Neumark}(1965)}]{Neumark}
\bibinfo{author}{\bibfnamefont{S.}~\bibnamefont{Neumark}},
  \emph{\bibinfo{title}{{Solution of Cubic and Quartic Equations}}}
  (\bibinfo{publisher}{Pergamon Press}, \bibinfo{year}{1965}).

\bibitem[{\citenamefont{Landau and Lifschitz}(1956)}]{Landau}
\bibinfo{author}{\bibfnamefont{L.~D.} \bibnamefont{Landau}} \bibnamefont{and}
  \bibinfo{author}{\bibfnamefont{E.~M.} \bibnamefont{Lifschitz}},
  \emph{\bibinfo{title}{{Mechanics}}}, vol.~\bibinfo{volume}{1} of
  \emph{\bibinfo{series}{Course of Theoretical Physics}}
  (\bibinfo{publisher}{Pergamon Press}, \bibinfo{year}{1956}).

\bibitem[{\citenamefont{Pi\~{n}a}(1997)}]{Pinna2}
\bibinfo{author}{\bibfnamefont{E.}~\bibnamefont{Pi\~{n}a}},
  \bibinfo{journal}{Rev. Mex. de F\'{i}s.} \textbf{\bibinfo{volume}{43}},
  \bibinfo{pages}{205} (\bibinfo{year}{1997}).

\bibitem[{\citenamefont{Pi\~{n}a}(2008)}]{PINAGARZA2008}
\bibinfo{author}{\bibfnamefont{E.}~\bibnamefont{Pi\~{n}a}},
  \bibinfo{journal}{{Rev. Mex. de F\'{i}s. E}} \textbf{\bibinfo{volume}{54}},
  \bibinfo{pages}{92 } (\bibinfo{year}{2008}).

\bibitem[{\citenamefont{Akhiezer}(1970)}]{Akhiezer}
\bibinfo{author}{\bibnamefont{Akhiezer}}, \emph{\bibinfo{title}{{Elements of
  the theory of elliptic functions}}} (\bibinfo{publisher}{American
  Mathematical Society}, \bibinfo{year}{1970}).

\bibitem[{\citenamefont{Sugiura}(1990)}]{Sugiura}
\bibinfo{author}{\bibfnamefont{M.}~\bibnamefont{Sugiura}},
  \emph{\bibinfo{title}{{Unitary Represntations and Harmonic Analysis: An
  Introduction}}} (\bibinfo{year}{1990}).

\bibitem[{\citenamefont{Brizard}(2013)}]{Brizard2013511}
\bibinfo{author}{\bibfnamefont{A.~J.} \bibnamefont{Brizard}},
  \bibinfo{journal}{Communications in Nonlinear Science and Numerical
  Simulation} \textbf{\bibinfo{volume}{18}}, \bibinfo{pages}{511 }
  (\bibinfo{year}{2013}).

\bibitem[{\citenamefont{Linares}(2016)}]{Linares}
\bibinfo{author}{\bibfnamefont{R.}~\bibnamefont{Linares}}
  (\bibinfo{year}{2016}), \eprint{1601.07891}.

\bibitem[{\citenamefont{Yoneya}(2016{\natexlab{b}})}]{Yoneya:2016abn}
\bibinfo{author}{\bibfnamefont{T.}~\bibnamefont{Yoneya}}
  (\bibinfo{year}{2016}{\natexlab{b}}), \eprint{1612.08509}.

\end{thebibliography}
\end{document}